\documentclass[amsmath,twocolumn,showpacs,amssymb,aps,nofootinbib]{revtex4-1}
\usepackage{graphicx}
\usepackage{subcaption}
\usepackage{pgfplots}
\usepackage{dcolumn}
\usepackage{bm}
\usepackage{slashed}
\usepackage{amsmath}
\usepackage{amssymb}
\usepackage{tikz}
\usepackage{bbold}
\usepackage{mathtools}
\usepackage{romannum}
\usepackage{natbib}
\usepackage{braket}
\begin{document}
\pagenumbering{arabic}
\title{Complexity and information geometry in spin chains\\}
\author{Nitesh Jaiswal}
\email{nitesh@iitk.ac.in}	
\author{Mamta Gautam}
\email{mamtag@iitk.ac.in}
\author{Tapobrata Sarkar}
\email{tapo@iitk.ac.in}
\affiliation{
Department of Physics, Indian Institute of Technology Kanpur-208016, India}
\date{\today}
\begin{abstract}
We study Nielsen complexity and Fubini-Study complexity for a class of exactly solvable one dimensional spin systems. 
Our examples include the transverse XY spin chain and its natural extensions, the
quantum compass model with and without an external magnetic field. 
We obtain the scaling behaviour of both complexities near quantum phase transitions in the thermodynamic limit, 
as a function of the system parameters. We provide 
analytical proofs of these, in an information geometric framework, which verify our numerical analysis. 
The scaling of the Nielsen complexity with the system size is also established, close to criticality. We also obtain 
analytic expressions for the Fubini-Study complexity in some special cases for
all the models, while a numerical analysis in more generic situations is carried out. Our study clearly 
demonstrates the differences in the two notions of complexity in quasi-free fermionic systems. 

\end{abstract}
\maketitle

\section{\label{Intro}Introduction}

Geometry plays a fundamental role in our understanding of physical phenomena at varied scales, from 
astrophysical scenarios to condensed matter systems. Although the origin of these phenomena might
be widely different, the geometric methods involved are largely similar. 
Indeed, Riemannian geometry lies at the core of much of modern physics, and can be successfully used to
address phenomena at extreme length scales, from general relativity to quantum mechanics. 

One of the major applications of Riemannian geometry in quantum statistical systems has been in the area of phase transitions. 
Phase transitions in physical systems continue to be the focus of intense research over the last many decades. 
In any system, a phase transition indicates a change from its original state due to variations in external 
control parameters, such as temperature, pressure, magnetic field, anisotropy, etc., with a possible symmetry breaking. 
Transitions involving a variation of 
the temperature are classical (thermal) phase transitions and are driven by thermal fluctuations, whereas 
those that occur at zero temperature are driven by quantum fluctuations and called quantum phase 
transitions (QPTs) \cite{Subir,Nigel, Cardy, Stanley}, and are extremely important in studying the low temperature 
behaviour of many statistical and condensed matter systems. 

Information theoretic geometry is known to provide 
valuable insights into the physics of both classical and quantum phase transitions (see, e.g., \cite{BrodyHook}). 
The basic idea arises from statistical systems, and dates back to early attempts at differentiating between
probability distributions. In the late 70's, this idea was extended to the distribution of classical particles 
via their statistical entropy, in what is now popularly known as thermodynamic geometry \cite{Ruppeiner}. 
Geometrical notions in quantum systems first arose in the work of \cite{Provost},
and has of late attracted strong interest, as these often provide useful and alternative indications of quantum
phase transitions. The essential procedure here is to establish the geometry of the parameter space, i.e., the
space of tunable parameters alluded to above. Indeed, there are several quantities associated with 
such a geometry, which can be used to study phase transitions. 

First, the quantum information metric (QIM) formulated in 
\cite{Provost} is the Riemannian metric induced on the parameter space of the Hamiltonian, and 
measures the ``distance'' between two neighboring quantum states. Such metrics are known to 
possess interesting scaling behaviour and are often singular, with the singular limits indicating rich physics.
Whereas one might argue that such singularities in metric components 
might be artifacts of specific coordinate choices, scalar quantities
which are independent of such choices, such as the Ricci scalar, often
show remarkable properties near QPTs \cite{Polkovnikov, Zanardi, Sarkar}. For example, this may diverge,
or show a discontinuity at a critical point (or, in general, along a critical line)
in quantum many-body systems, although contrary results
have also been reported in the literature \cite{TapoDickeModel}. We also recall at this point that 
the QIM of the ground state of a many body system is the real part of a more general structure -- the quantum 
geometric tensor (QGT) -- whose imaginary part is the celebrated Berry phase. 

In this paper, we will be interested in a new information theoretic quantity known as complexity, that has, of late,
become extremely popular in the literature. The concept of complexity is by now ubiquitous in physics, although
traditionally it appeared in branches of mathematics related to computer science.
From the perspective of quantum mechanics, complexity quantifies the difficulty in constructing 
a unitary transformation to reach a desired target state from a given reference state. 
In other words, the complexity of a target state is defined as the number of quantum operations called 
gates, built in an optimal circuit, to arrive at the target state from the original reference state. 
This procedure is called ``gate counting,'' and complexity measured in this way is called ``circuit complexity.'' 
A geometric method was proposed by Nielsen \cite{Nielsen} (for recent applications, 
see e.g. \cite{Liu, Khan, Xiong}) to find this optimal circuit, and we
will refer to this as the Nielsen complexity (NC). NC has been particularly well studied in the 
high energy physics literature. There, the excitement has largely been in the context of holography, i.e., 
the gauge-gravity duality, and of late, the issue of complexity has been revived in quantum field theory 
(where explicit computations are difficult per se), with the idea being that 
this can be understood via a holographic computation (see, e.g., \cite{Myers1, Myers2, Diptarka, Arpan, Arpan2, Guo},
and references therein for some recent work on this aspect).	

On the other hand, information theoretic geometry 
has been recently used for an alternative construction of the complexity, in \cite{Chapman}.  
In that work, the complexity of a quantum field theory was defined in terms of a minimal (geodesic) length as
determined by the Fubini-Study metric on the space of Gaussian states. We will call this the
Fubini-Study complexity (FSC). The purpose of this paper is to study the NC and the FSC for some exactly solvable
many body quantum statistical systems, for which the metric on the parameter space is the QIM 
that we have mentioned before. 

In this work, we study the NC and the FSC for three different exactly solvable one dimensional 
spin systems : the transverse spin-$1/2$ XY model, and the quantum compass models with and without a
transverse magnetic field. Our main results are as follows. 
First, using information geometry, we establish a relation between the infinitesimal NC and the FSC. 
Next, for the NC, we establish its non-analyticity near a QPT, and find 
a finite size scaling relation of its derivative across such a transition. We also obtain the scaling behaviour 
of the derivative of the NC with the system 
parameters, close to a QPT, in the thermodynamic limit. We give an analytical proof of the latter. 
We find that the behaviour of the derivative of the NC is similar to that of the derivative of the 
Berry phase, close to a QPT. Further, we establish that the NC might be
a better indicator of QPTs compared to the Ricci scalar, for Ising transitions. 

Next, in the context of the FSC, we obtain analytical expressions for the geodesic length in 
some special cases, with simplifying assumptions. These show that the FSC is also non-analytic at a 
QPT, where its derivative diverges. We establish the scaling behaviour of the 
derivative of the FSC with the system parameters near a QPT, by a mathematical argument. 
Next, we consider generic geodesics that we numerically solve, and confirm this scaling.  

The organization of this paper is as follows. In Sec. \ref{complexity}, we 
set up the basic notations and conventions. We show here how the infinitesimal Nielsen complexity
is related to the quantum information metric. Section \ref{transversexy} studies our first
example, the transverse XY spin-$1/2$ chain. In section \ref{compass}, a similar analysis for
the quantum compass model is reported, and this is extended in section \ref{compassModH} for the
compass model in a transverse magnetic field. Finally, the paper ends with some discussions and conclusions in
section \ref{Conclusions}. Two appendices are included in this paper, which show some mathematical details of the 
computations for the XY model and for the compass model in a magnetic field. 

\section{\label{complexity} The information metric and complexity}

The QIM defines the distance between two nearby quantum states \(|\psi(\vec{\lambda})\rangle\) 
and \(|\psi(\vec{\lambda}+d\vec{\lambda})\rangle\), characterized by 
a set of parameters ${\vec \lambda}$ and separated by an infinitesimal amount 
\(d\vec{\lambda}\) in the parameter space,
\begin{eqnarray}
d\tau^{2}&=&1-|\langle\psi(\vec{\lambda})|\psi(\vec{\lambda}+d\vec{\lambda})\rangle|^{2} \notag\\
&=&g_{ij}d\lambda^{i}d\lambda^{j}+	{\cal O }(|d\vec{\lambda}|^{3})~,
\label{genline}
\end{eqnarray}
where the quantum information metric \(g_{ij}\) is the real symmetric part of the QGT,
denoted by \(\chi_{ij}\) and a summation is implied over repeated indices. 
The expression for the QGT is 
\begin{equation}
\chi_{ij}=\langle\partial_{i}\psi|\partial_{j}\psi\rangle-\langle
\partial_{i}\psi|\psi\rangle\langle\psi|\partial_{j}\psi\rangle~,
\label{qgt}
\end{equation}
with \(\partial_{i}\equiv\frac{\partial}{\partial\lambda^{i}}\), \(i=1,2,\cdots,m\), 
where \(m\) is the dimension of the parameter space, and that for the metric reads
\begin{equation}
g_{ij}=Re[\chi_{ij}]=\frac{1}{2}\left(\chi_{ij}+\chi_{ji}\right)~.
\end{equation}
Here, we will be mostly interested in two-dimensional parameter spaces for which $m=2$, and for these,
once the metric is computed, and if it is diagonal, we can calculate the scalar curvature $R$, from
\begin{equation}
R = \frac{1}{\sqrt{g}}\left[\frac{\partial}{\partial \lambda^1}\left(\frac{1}{\sqrt{g}} 
\frac{\partial g_{22}}{\partial \lambda^1}\right)_{\lambda^2} + 
\frac{\partial}{\partial \lambda^2}\left(\frac{1}{\sqrt{g}} \frac{\partial g_{11}}{\partial \lambda^2}\right)_{\lambda^1}\right]~,
\label{scalarcurvature}
\end{equation}
with $g$ being the determinant of the metric. 
There is a slightly more complicated formula for $R$, for non-diagonal metrics, which we omit here for brevity. 
For these two-dimensional parameter manifolds (corresponding to equilibrium ground states of spin systems)
that we will be interested in here, $R$ uniquely specifies the curvature properties,
and is an important quantity to study. As is well known, $R$ possesses some interesting properties
at quantum phase transitions -- it might diverge there, or might show a discontinuity, although as we have
mentioned, exceptions to such behaviour are also documented in the literature \cite{TapoDickeModel}. 

Important to us will be the Bogoliubov angle $\theta_k$ that characterizes the ground state of a diagonalized
quadratic Hamiltonian, and depends on the system parameters ${\vec \lambda}$, with $k$ being the Fourier index. 
Suppose we have a two-parameter model as above, then we can write an infinitesimal 
change of the Bogoliubov angle (from a given reference state) as 
\begin{equation}
d\theta_k = \left(\frac{\partial\theta_k}{\partial\lambda^1}\right)d\lambda^1 + 
\left(\frac{\partial\theta_k}{\partial\lambda^2}\right)d\lambda^2~,
\end{equation}
where the derivatives are evaluated at the ground state. Now if we divide throughout by $2$, 
square this expression and sum over all the momentum modes, then we obtain
\begin{eqnarray}
&~&\frac{1}{4}\sum\limits_{k}^{}\left(d\theta_k\right)^2 =\frac{1}{4}\sum\limits_{k}^{}\left(\frac{\partial\theta_k}{\partial\lambda^1}
\right)^2(d\lambda^1)^2 \nonumber\\
&+& \frac{1}{2}\sum\limits_{k}^{}\left(\frac{\partial\theta_k}{\partial\lambda^1}
\right)\left(\frac{\partial\theta_k}{\partial\lambda^2}\right)(d\lambda^1d\lambda^2)
+ \frac{1}{4}\sum\limits_{k}^{}\left(\frac{\partial\theta_k}{\partial\lambda^2}
\right)^2(d\lambda^2)^2~.\nonumber\\
\label{FSC}
\end{eqnarray}
For quasi-free fermionic systems with quadratic Hamiltonians, the right hand side is the line element in the parameter space
of eq.(\ref{genline}), 
$d\tau^2=g_{ij}d\lambda^i d\lambda^j$, with the QIM defined in terms of the Bogoliubov angle being \cite{Zanardi},
\begin{equation}
g_{ij} = \frac{1}{4}\sum\limits_{k}^{}\left(\frac{\partial\theta_k}{\partial\lambda^i}\right)~
\left(\frac{\partial\theta_k}{\partial\lambda^j}\right)~.
\label{gij}
\end{equation}
The components of the QIM can often be computed explicitly in the thermodynamic limit where the summation 
is replaced by an integration, i.e., \(\sum\limits_{k}^{}\longrightarrow\frac{N}{2\pi}\int_0^\pi dk\). \footnote{In
what follows, the complexity, metric, scalar curvature etc. will be calculated per site $N$. Unless required 
otherwise, we will not mention this further in their formulae, in order not to clutter the notation.}
The integrals are in turn calculated by a standard process of residue evaluation on the complex plane. 

Now, for the quadratic Hamiltonians that we are interested in, the NC (which will henceforth be denoted by
the symbol ${\mathcal C}_{N}$) can be shown to be given by (this was shown
in \cite{Liu}, and for completeness, we will illustrate this formula for one of our examples in appendix \ref{AppendixA})
\begin{equation}
{\mathcal C}_N =\sum\limits_{k}^{}|\Delta\theta_k|^2~,
\label{NC}
\end{equation}
which is the square of the (finite) difference of the Bogoliubov angles between an initial (reference) state $\ket{\Psi_R}$
and a final (target) state $\ket{\Psi_T}$, with $\Delta\theta_k = (\theta_k^T - \theta_k^R)/2$.
This is obtained by considering a trajectory in the space of unitary transformations that minimises 
the cost (or length) functional in such a space. Importantly, eq.(\ref{NC}) is valid only for the so called 
$\kappa = 2$ cost functions (see discussion after eq.(\ref{CostFunction}) in appendix \ref{AppendixA}), and to which
we will restrict ourselves here. 
For other cost functions, the geometry mentioned above is not Finsler, and hence the notion of a
distance is not well defined, although these cases can be well approximated by Finsler geometries \cite{Nielsen}.
A discussion on general cost functions appear in \cite{Myers1}, \cite{Myers2}, and the authors of \cite{Myers3} 
analyzed various cost functions for free fermionic theories. 

The FSC (to be henceforth denoted by the
symbol ${\mathcal C}_{FS}$) on the other hand, can be conveniently
defined as an integral over the infinitesimal differences quantified in eq.(\ref{FSC}), over a given 
trajectory in the parameter space. Formally, we have for the specific class of cost functions 
discussed above, 
\begin{equation}
{\mathcal C}_{FS} = \int_{\mathcal P} \sqrt{d{\mathcal C}_N} = \int_{\mathcal P} d\tau~,
\label{Cin}
\end{equation}
with $d{\mathcal C}_N$ defined by the left hand side of eq.(\ref{FSC}) (and is not to be interpreted 
as the differential of the right hand side of eq.(\ref{NC})), and ${\mathcal P}$ is a specified
path. The FSC is thus given by an integral over the square root of the NC of infinitesimally separated states 
that are related by a unitary transformation. One has to now specify the path ${\mathcal P}$. A 
natural choice is a path of minimum length between two given points on the parameter manifold, 
called a geodesic path. This is then the final prescription for computing the FSC for equilibrium
ground states of quadratic
Hamiltonians. Starting from a given point on the manifold, one reaches an infinitesimally
separated neighbouring point by an optimal operation in the unitary space, and then
integrates the differential path length along a geodesic in the parameter space. Since $d{\mathcal C}_N$ is related
to the QIM, one can then try to find analytic expressions for the right hand side of eq.(\ref{Cin}),
once the geodesic path on the parameter manifold is specified. 

To make this more concrete, let us briefly recall a few elementary facts about geodesics. 
For a Riemannian manifold with coordinates $\lambda^i$ ($i=1,2$ for our purposes) 
that is equipped with a metric $g_{ij}$, a geodesic is a path that extremizes the proper 
distance (or the square root of the line element). 
This can be cast as a variational problem, to determine the extrema of 
the integral $\int_1^2 \sqrt{g_{ij}{\dot \lambda^{i}}{\dot \lambda^{j}}}d\alpha$,
where the dot denotes a derivative with respect to $\alpha$, which is an affine parameter that
parametrizes a curve in the two dimensional manifold that joins the two points $1$ and $2$. 
Calculus of variations then reveals that geodesic curves are solutions to the differential equations 
(with $\Gamma$ denoting the Christoffel connection) 
\begin{equation}
{\ddot \lambda^{i}} + \Gamma^{i}_{jk}{\dot \lambda^{j}}{\dot \lambda^{k}} = 0~,
~\Gamma^{i}_{jk} = \frac{1}{2}g^{im}\left(\frac{\partial g_{mj}}{\partial \lambda^{k}} + \frac{\partial g_{mk}}{\partial \lambda^{j}}
- \frac{\partial g_{jk}}{\partial \lambda^{m}}\right)~.
\label{geodesic}
\end{equation}
The above can also be obtained by extremizing the Lagrangian
\begin{equation}
{\mathcal L} = \frac{1}{2}\sqrt{g_{ij}{\dot \lambda^{i}}{\dot \lambda^{j}}}
\label{Lagrangian}
\end{equation}
and using the Euler-Lagrange equations that follow. Importantly for our purpose, 
it is useful to consider geodesics parametrized
by $\tau$ itself. In that case, we have that the vector 
$u^i={\dot \lambda^{i}} = d\lambda^{i}/d\tau$ is normalized, such that $u^iu_i=$ 
${\dot \lambda^{i}}{\dot \lambda_{i}} = g_{ij}{\dot \lambda^{i}}
{\dot \lambda^{j}} = 1$ (a condition which is imposed after the extremization). 

In general, eq.(\ref{geodesic}) provides a set of coupled non-linear differential equations, which might be analytically
intractable, and one has to resort to a numerical analysis. 
However, computations can be simplified if one can find a cyclic coordinate \cite{TapoKumar}. 
In that case, we can use the Euler-Lagrange
equation corresponding to the cyclic coordinate, in conjunction with the normalization condition, to obtain
a set of simultaneous algebraic equations that can be solved to yield ${\dot\lambda^i}$. The resulting 
set of first order differential equations can then be used to obtain a solution for the non-cyclic coordinate 
$\lambda^j(\tau)$. This will yield a solution with one arbitrary constant that is fixed by choosing a convenient starting
point, and the resulting solution can be inverted to 
obtain $\tau(\lambda^j)$. Since ${\mathcal C}_{FS} = \tau$, i.e. 
the length of the geodesic (measured from the chosen starting point), this will yield the FSC in terms of
the system parameters, and in particular can be useful in understanding the behaviour of the FSC near
QPTs. 

Of course, such cyclic coordinates are hard to come by (we will show a special case in which it exists in the parameter space, 
for the transverse XY model in the next section). Its existence is however 
guaranteed if we consider Hamiltonians that are phase-rotated by 
an angle about a given axis. The rotation angle will play no role in the spectrum of the model, since it enters the ground
state as a phase, but this extra degree of freedom will conveniently give the cyclic coordinate that we are looking for. 
The parameter space now becomes three dimensional with a non-trivial metric component along the cyclic coordinate, 
and we will restrict to two-dimensional planes with one of the coordinates being the rotation angle. As is known, 
information geometry restricted to these planes capture the essential features of the phase diagram of the theory, and hence
are useful to study in the context of complexity. 

Note also that the NC defined from eq.(\ref{NC}) or equivalently from the left hand
side of eq.(\ref{FSC}) does not depend on this phase angle (since the Bogoliubov angle does not). 
From the definition of eq.(\ref{Cin}) therefore,
our geodesics should be lines with constant values of the cyclic coordinate. These give useful 
results for the FSC since these are the geodesics which reach the singular points of the parameter manifold, as we will see.  

For more general geodesics, we will resort to an appropriate numerical analysis of eq.(\ref{geodesic}), 
and this will be elaborated later in this paper.

\section{\label{transversexy} The transverse XY model}

We will consider the one-dimensional spin-$1/2$ XY model in a transverse magnetic field. This is one
of the most well studied models in statistical mechanics that exhibits quantum phase transitions
\cite{LSM, Katsura1, BarouchMcCoy, Bunder, Amit} (see also the excellent recent monograph \cite{AmitBook}). 
We start with the Hamiltonian
\begin{equation}
H=-\!\!\sum\limits_{l=-M}^{M}\!\!\left(\frac{1+\gamma}{4}\sigma^{x}_{l}\sigma^{x}_{l+1}
+\frac{1-\gamma}{4}\sigma^{y}_{l}\sigma^{y}_{l+1}+\frac{h}{2}\sigma^{z}_{l}\right),
\end{equation}
where $M=(N-1)/2$, for odd $N$, $\gamma$ is an anisotropy parameter, and $h$ the applied magnetic field. 
Also, $\sigma$ denotes the Pauli matrices. As we mentioned towards the end of the last section, 
we consider a modified Hamiltonian, obtained by applying a rotation of $\phi$ around the $z$-direction,
\begin{equation}
H(\phi)=g(\phi)	H g^{\dagger}(\phi)\quad	\text{with}\quad g(\phi)=\!\!\prod\limits_{l=-M}^{M}\!\!\exp
\left(i\sigma^{z}_{l}\frac{\phi}{2}\right)~,
\label{rotHamiltonian}
\end{equation}
After a standard usage of the Jordan-Wigner, Fourier and Bogoliubov transformations, 
the eigenvalues of the Hamiltonian of eq.(\ref{rotHamiltonian}) can be written as \cite{AmitBook}
\begin{equation}
\Lambda_{k\pm}=\pm\sqrt{(\cos k-h)^{2}+(\gamma\sin{k})^{2}},
\end{equation}
where $k$ takes the values
\begin{equation}
k=\frac{2\pi\lambda}{N}~,~~\lambda=-\frac{N-1}{2},....,-1,0,1,....,\frac{N-1}{2}~.
\label{krange}
\end{equation}
From the single particle energy spectrum, the energy gap is calculated to be 
\begin{eqnarray}
\Delta(k)=\Lambda_{k+}-\Lambda_{k-}
=2\sqrt{(\cos k-h)^{2}+(\gamma\sin{k})^{2}}~.\nonumber\\
\label{energygap}
\end{eqnarray}
QPTs occur where the spectrum is gapless in the parameter space. Here, it happens 
on the line $\gamma=0,~|h| \leq 1$, which is the anisotropic transition line between two ferromagnetically 
ordered phases of the model, and at $|h| = 1$ (for $k=0, \pi$), which are the Ising transition lines between a ferromagnetic and 
a paramagnetic phase, with a spontaneously broken ${\mathbb Z}_2$ symmetry. 
The ground-state of the rotated Hamiltonian of eq.(\ref{rotHamiltonian}) is
\begin{eqnarray}
\ket{g}&=&\prod\limits_{k=0}^{\pi}\bigg[\cos\left(\frac{\theta_{k}}{2}\right)\ket{0}_{k}\ket{0}_{-k}
-i\sin\left(\frac{\theta_{k}}{2}\right)\times\notag\\
& &\quad \quad \:e^{2i\phi}\ket{1}_{k}\ket{1}_{-k}\bigg]~,
\label{gsXY}
\end{eqnarray}
where $\ket{0}_k$ and $\ket{1}_k$ denote the vacuum and the single excitation states of Jordan-Wigner
fermions with momentum $k$, and the Bogoliubov angle $\theta_k$ is defined from
\begin{equation}
\cos\theta_{k}=\frac{\cos k-h}{\sqrt{(\cos k-h)^{2}+(\gamma\sin{k})^{2}}}~.
\label{Bogoliubov}
\end{equation}
Following \cite{Liu}, the NC can be computed straightforwardly for this model. We have relegated the
details to the appendix, and the final form of the complexity appears in eq.(\ref{genericParaj}), which
was also written in eq.(\ref{NC}). Note that 
eq.(\ref{NC}) is a generic formula for quadratic Hamiltonians as demonstrated in \cite{Liu}, and will be true for 
all examples considered here. We will not provide the details for our other models to follow.
\begin{figure}
\centering
\begin{subfigure}{0.48\columnwidth}
\includegraphics[width=\textwidth]{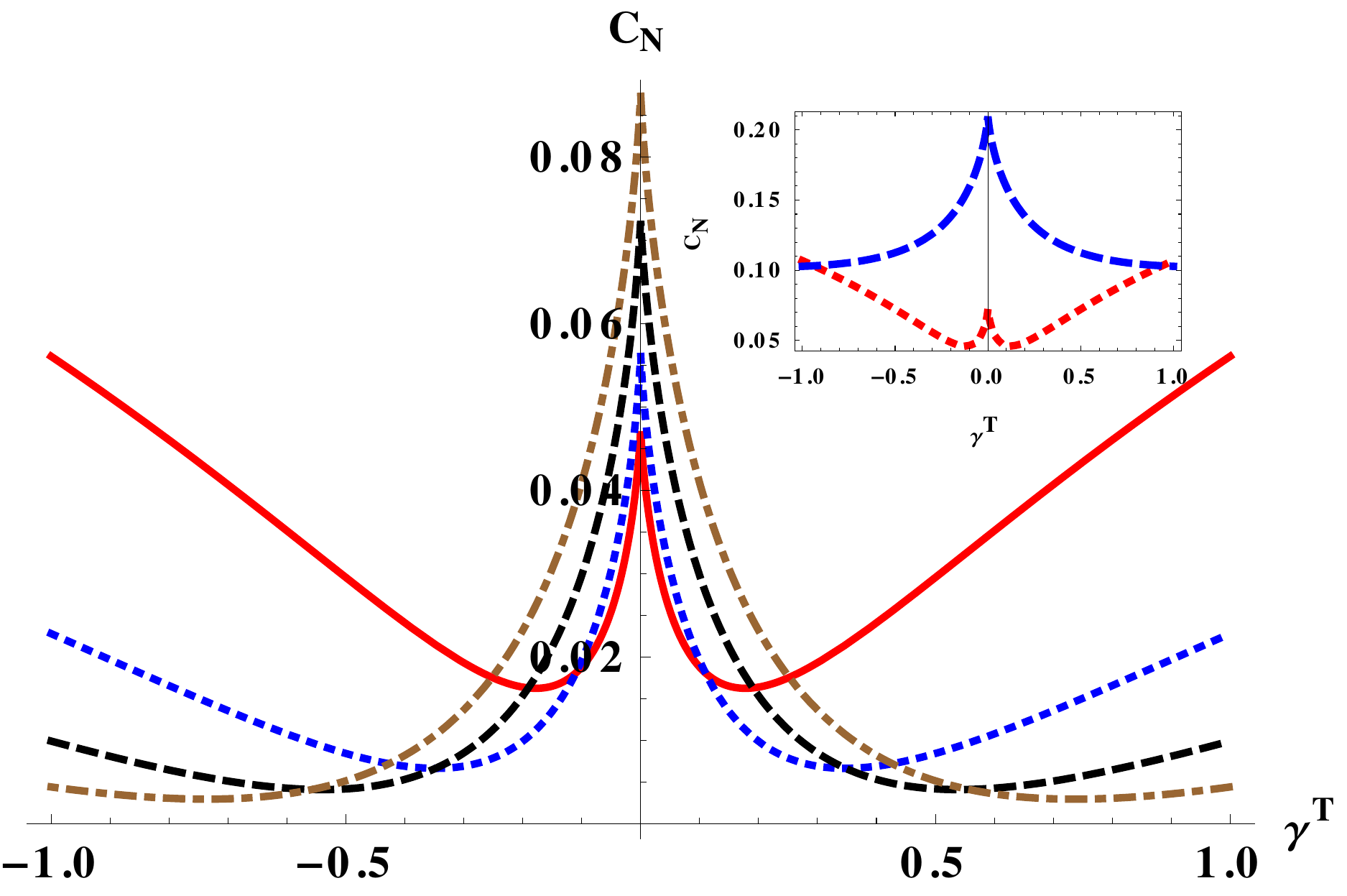}
\caption{NC of the transverse XY model as a function of $\gamma^T$}
\end{subfigure}
\hfill
\begin{subfigure}{0.48\columnwidth}
\includegraphics[width=\textwidth]{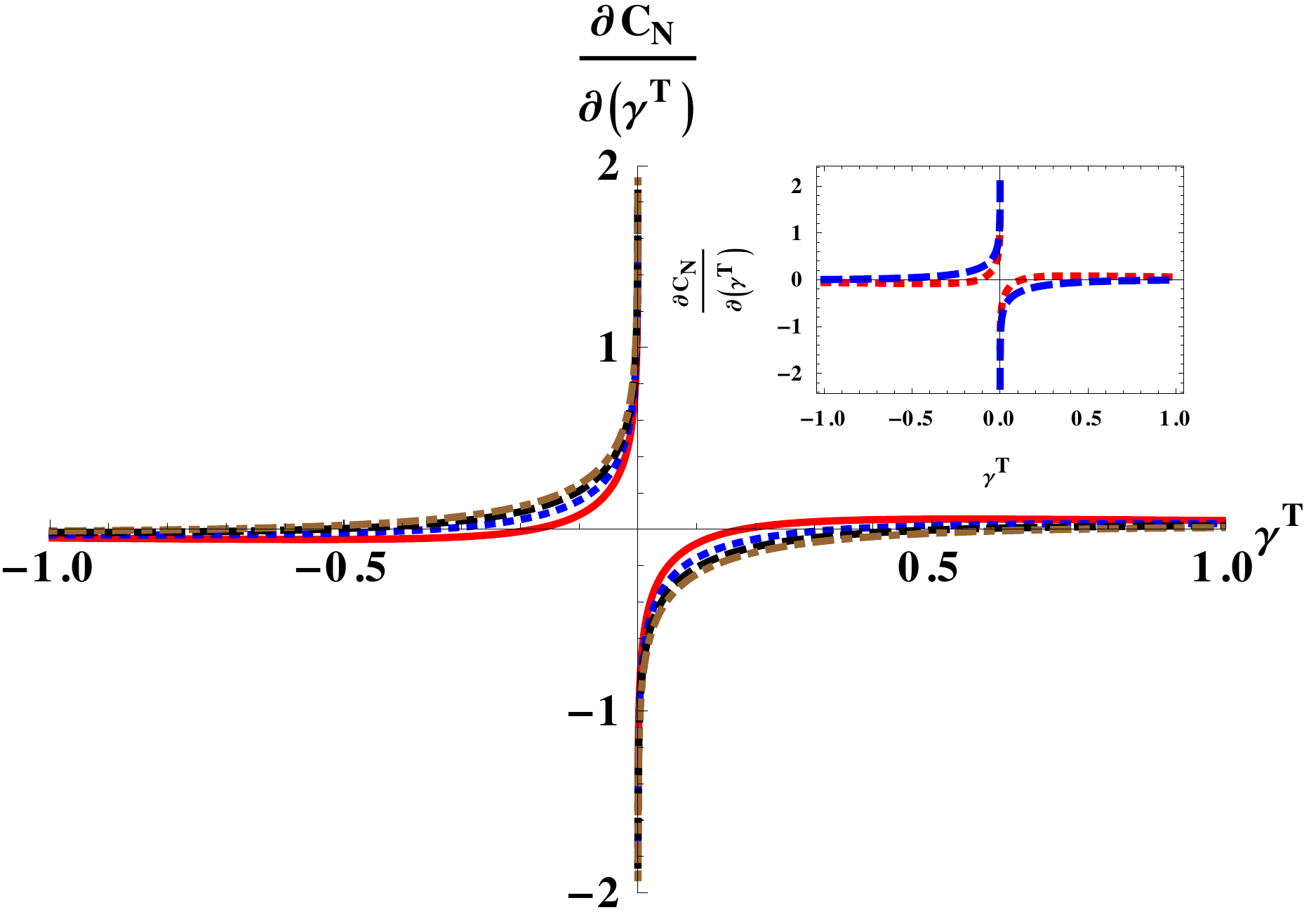}
\caption{Derivative of the NC with respect to $\gamma^T$} 
\end{subfigure} 
\caption{The NC (a) and its derivative (b) as a function of $\gamma^T$, for the transverse XY model. 
The insets correspond to reference states on a critical line (see text).} 
\label{figxymodel0}
\hfill
\begin{subfigure}{0.48\columnwidth}
\includegraphics[width=\textwidth]{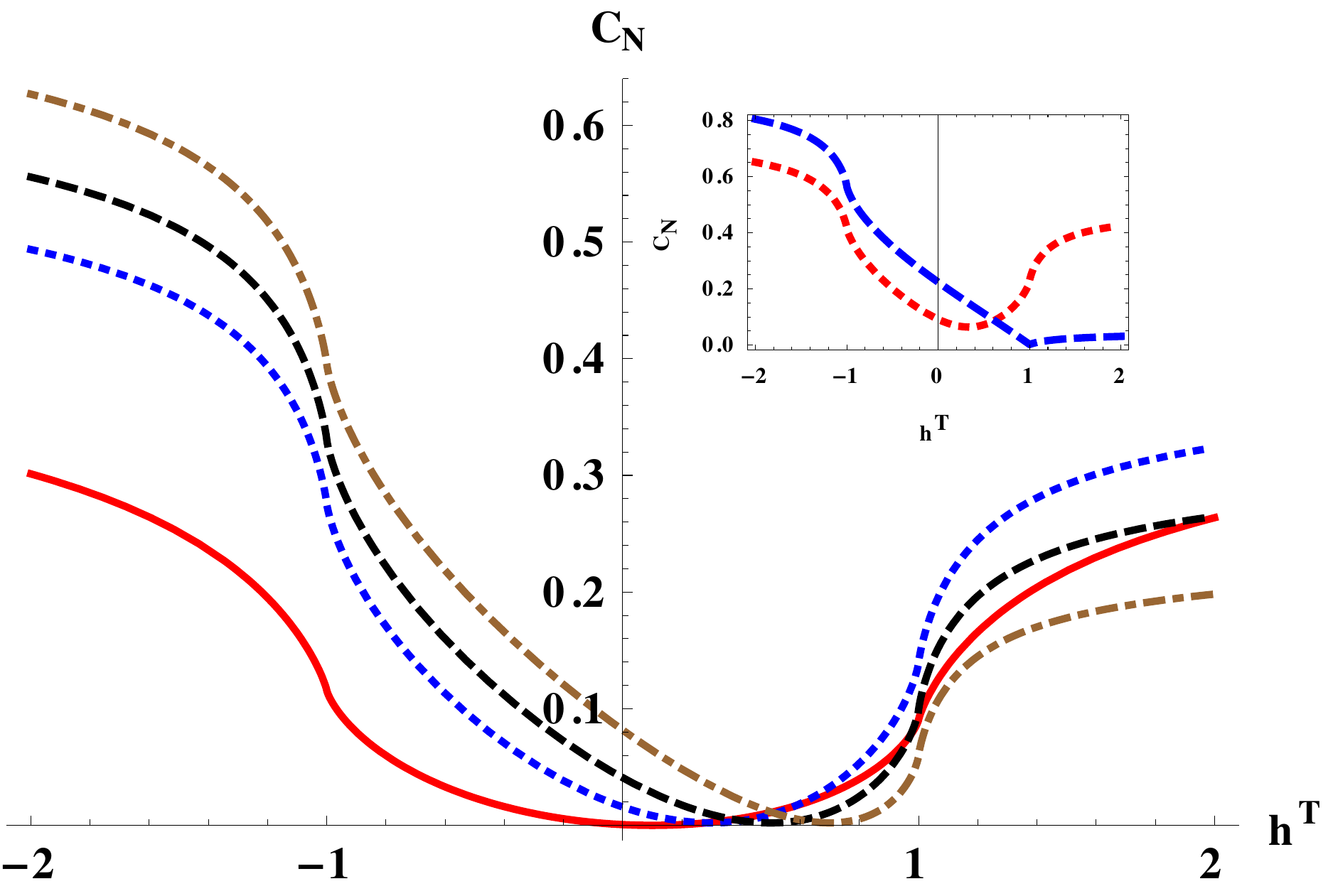}
\caption{NC of the transverse XY model as a function of $h^T$}
\end{subfigure}
\hfill
\begin{subfigure}{0.48\columnwidth}
\includegraphics[width=\textwidth]{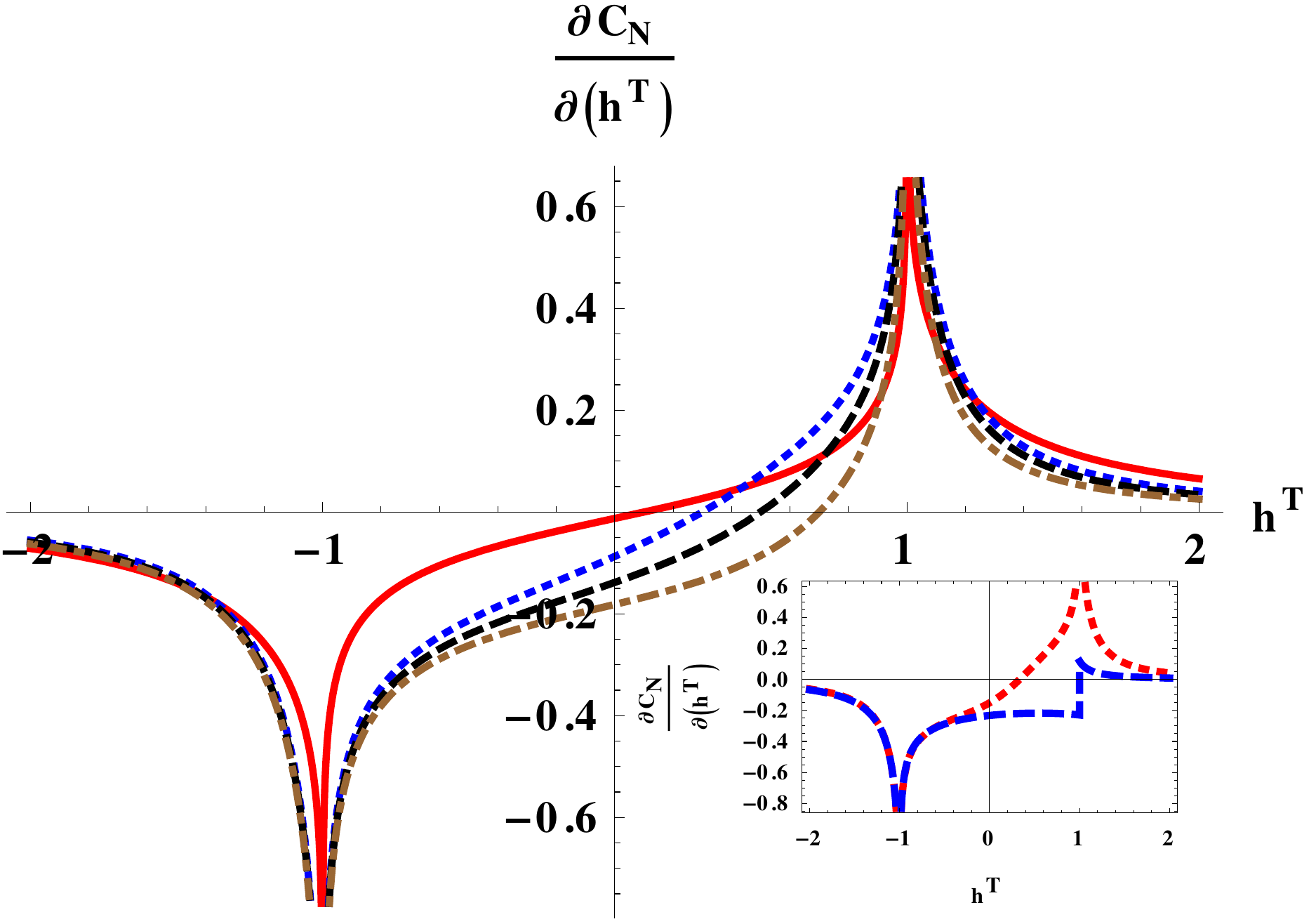}
\caption{Derivative of the NC with respect to $h^T$} 
\end{subfigure} 
\caption{The NC (a) and its derivative (b) as a function of $h^T$, for the transverse XY model. 
The inset corresponds to reference states on a critical line (see text).} 
\label{figxymodel}
\end{figure}  

In fig.(\ref{figxymodel0})(a), we show the NC and in fig.(\ref{figxymodel0})(b) 
its derivative with respect to the target parameter $\gamma^T$, for our transverse XY spin chain, 
in the thermodynamic limit. Here, in the main figures, 
we have taken for illustration, $h^R = 1/3, h^T=1/2$, and the solid red, dotted blue, dashed black and dot-dashed 
brown curves correspond to $\gamma^R=0.1$, $0.3$, $0.5$ and $0.7$, respectively. 
The insets show the situations when the reference state is on a 
critical line. The dotted red and dashed blue lines in the inset of fig.(\ref{figxymodel0})(a) show
the NC as a function of $\gamma^T$ for $(\gamma^R=0, h^R = 1/3, h^T=1/2)$ and 
$(\gamma^R=1/3, h^R = 1,h^T=1/2)$, respectively, with the same color coding in the inset
of fig.(\ref{figxymodel0})(b) which depicts their derivatives with respect to $\gamma^T$. 
The first of these sets of values correspond to a generic point on the anisotropy transition line 
$(\gamma^R = 0, |h^R|\leq 1)$, and the second to a generic point on the Ising transition line $h^R=1$.\footnote{Multi 
critical points, which occur at the intersection of the anisotropy transition line and the Ising transition
line will not be considered in this paper. }

In fig.(\ref{figxymodel})(a) and (b), we similarly depict the NC and its derivative respectively, with respect to the target 
parameter $h^T$. Here, in the main figures, the solid red line (which will be important for us in the
discussion of the next section) corresponds to $\gamma^R=\gamma^T=1$, with
$h^R=0.1$, and the values $\gamma^R = 1/3, \gamma^T = 1/2$ have been chosen for all other lines in the main 
figs.(\ref{figxymodel})(a) and (\ref{figxymodel})(b). The dotted blue, dashed black and dot-dashed brown curves
here correspond to $h^R= 0.3$, $0.5$ and $0.7$, respectively.

The insets of these figures depict the situation for reference states on critical lines. In particular,
the dotted red line corresponds to $(\gamma^R=0,h^R=1/3,\gamma^T=1/2)$ and the dashed blue 
line corresonds to $(\gamma^R=1/3,h^R=1,\gamma^T=1/2)$.
The reader might notice the non-divergent nature of latter curve in
the inset of fig.(\ref{figxymodel})(b). This will be explained momentarily. 

Clearly, when the reference state is away from criticality, and is taken through the quantum phase transition 
by changing the target state parameter $\gamma^T$ or $h^T$, the derivative of the NC diverge 
as $\gamma^T \to 0$ and $|h^T| \to 1$. 
This behaviour is similar to that of the Berry phase, which also shows similar 
non-analyticity across second order quantum phase transitions. In what follows, we will make this 
more precise. 

The behaviour of the derivative of the NC at a QPT is best explained from eq.(\ref{NC}), from which we glean that 
$\partial {\mathcal C}_N/\partial \gamma^T = \sum_k 2|\Delta\theta_k|({\rm sign}\Delta\theta_k)
(\partial\Delta\theta_k/\partial\gamma^T) =
\sum_k  2\Delta\theta_k(\partial\Delta\theta_k/\partial\gamma^T)$, and in 
a similar manner, it follows that $\partial {\mathcal C}_N/\partial h^T = \sum_k  2\Delta\theta_k
(\partial\Delta\theta_k/\partial h^T)$.
Now, from the definition of $\Delta\theta_k$ given after eq.(\ref{NC}), it is seen that 
\begin{eqnarray}
\frac{\partial\Delta\theta_k}{\partial \gamma^T} &=& \frac{1}{2}{\rm sign}(\gamma^T)\frac{\left(\cos k - h^T\right)\sin k}
{\left(h^T-\cos k\right)^2+ (\gamma^T)^2\sin^2 k}~,\nonumber\\
\frac{\partial\Delta\theta_k}{\partial h^T} &=& \frac{1}{2}{\rm sign}(\gamma^T)
\frac{\gamma^T\sin k}{\left(h^T-\cos k\right)^2+ (\gamma^T)^2\sin^2 k}~.~~\nonumber\\
\label{NCdiv}
\end{eqnarray}
These derivatives diverge at the quantum phase transition, as follows from eq.(\ref{energygap}). Of course, 
one has to also be careful about the behaviour of $\Delta\theta_k$, as this might vanish for some particular
values of $k$. 

To gain insight, we first look at the anisotropic transition line, with $\gamma^T=0$. 
As seen from the first relation of eq.(\ref{NCdiv}), in the limit $\gamma^T \to 0$, 
$\partial\Delta\theta_k/\partial \gamma^T \sim \sin k/(h^T - \cos k)$, i.e., diverges near the critical line 
as $1/k'$, with $k' = k - \arccos(h^T)$.
Now, we expand $\theta_k^T$ close to criticality, in powers of the 
shifted momentum $k'$. A simple exercise shows that in the limit
$\gamma^T\to 0$, we have $\theta_k^T = \arccos({\rm sign}(k'))$.
When one approaches the critical line, its limiting values are $\pi$ and $0$
as one takes the limit $k' \to 0$ from the right or the left, respectively. 
Then, if $\theta_k^R \neq \pi$ or $0$ near criticality (which will be generally true
for the cases considered here), the divergence of the derivative of the NC across the line
$\gamma^T = 0$ is due to the singularity at $k'=0$. There is a subtlety that arises when $\gamma^R=0$ (the dotted red line in 
the inset of fig.(\ref{figxymodel0})(a)). Here, since the initial state is already on the critical line, one has
to carefully construct the limiting value of $\Delta\theta_k$. We will not go to the details here, but state
the result that $\Delta\theta_{k}$ is finite at $k'=0$, resulting in the divergence of the derivative of the NC here,
as depicted in the inset of fig.(\ref{figxymodel0})(b). 

It is convenient, at this stage, to discuss the scaling of $\partial {\mathcal C}_N/\partial \gamma^T$ 
with the system size. Since the divergent part of 
$\Delta\theta_k\partial\Delta\theta_k/\partial \gamma^T \sim 1/k'$, it follows straightforwardly that  
$\partial {\mathcal C}_N/\partial \gamma^T
\sim N\log N$, with $N$ being the number of lattice sites in eq.(\ref{krange}).
\begin{figure}[h!]
\centering
\begin{subfigure}{0.48\columnwidth}
\includegraphics[width=\textwidth]{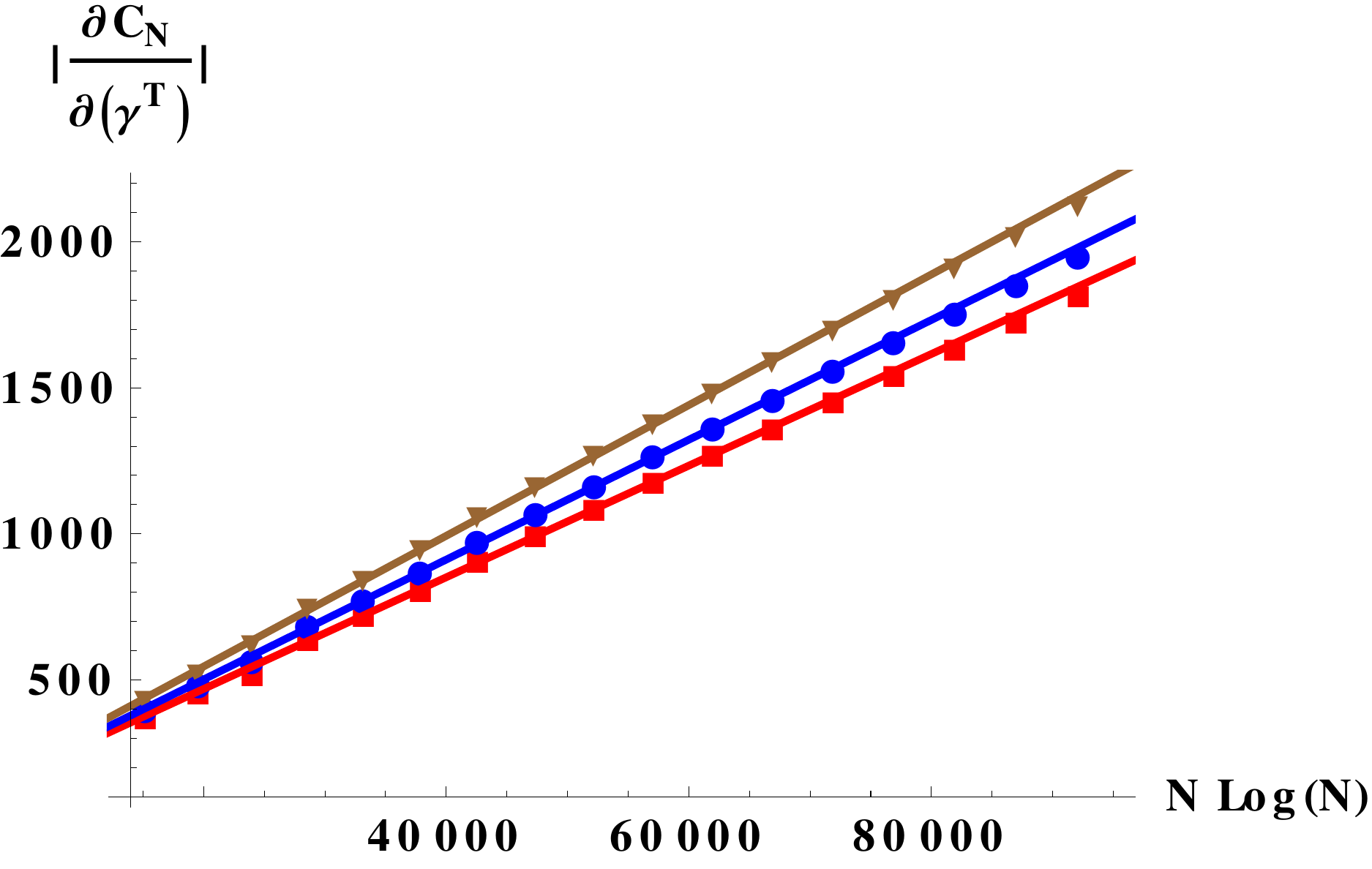}
\caption{Scaling of $\partial {\mathcal C}_N/\partial \gamma^T$ with the system size $N$.}
\end{subfigure}
\hfill
\begin{subfigure}{0.48\columnwidth}
\includegraphics[width=\textwidth]{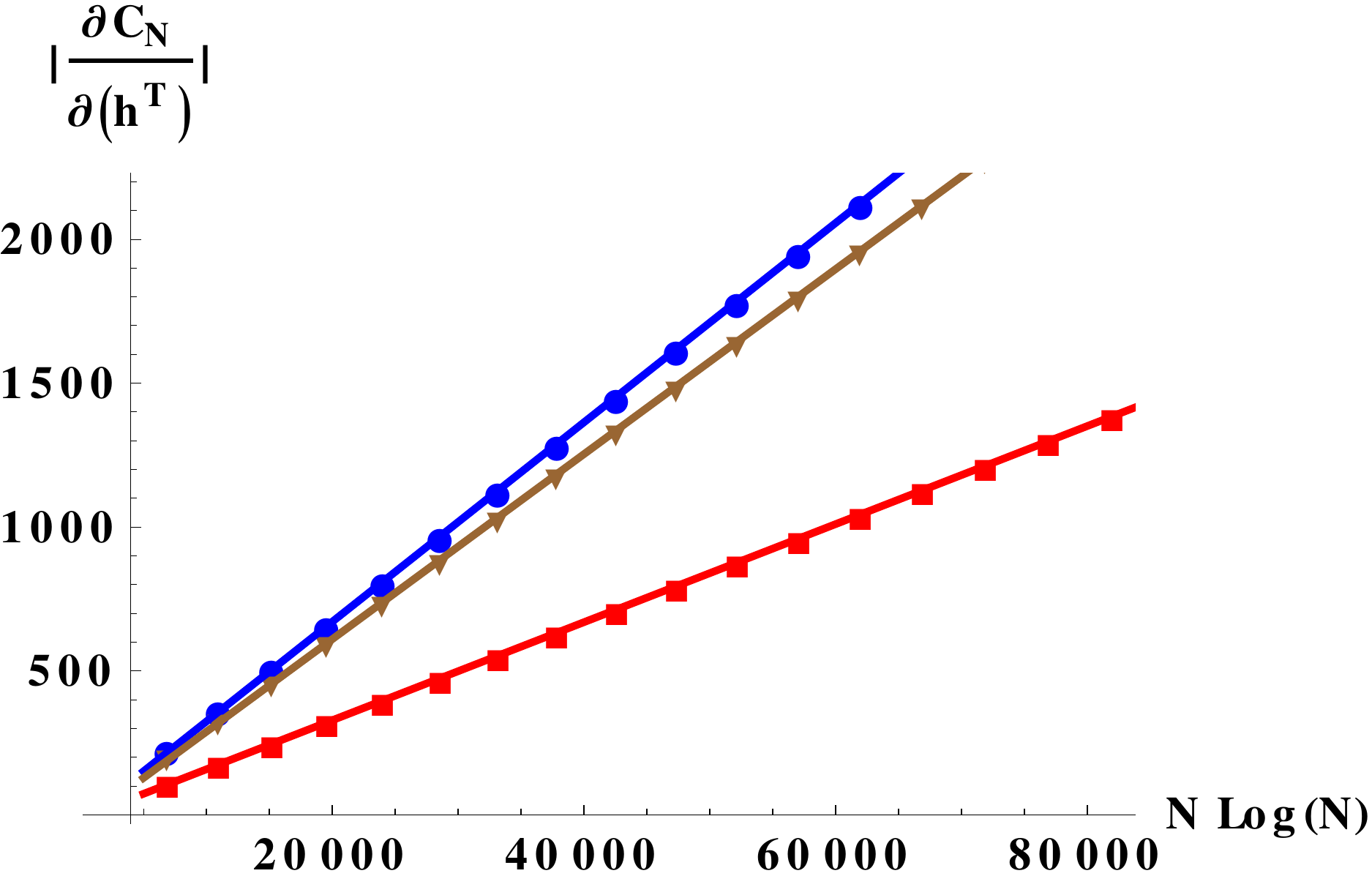}
\caption{Scaling of $\partial {\mathcal C}_N/\partial h^T$ with the system size $N$.} 
\end{subfigure} 
\caption{Scaling of the derivative $\partial {\mathcal C}_N/\partial \gamma^T$ (a) and 
$\partial {\mathcal C}_N/\partial h^T$ (b) as a function of the system size $N$, for the transverse XY model. 
} 
\label{scaling}
\end{figure}

To check this explicitly, we have, in fig.(\ref{scaling})(a), plotted the numerically obtained values 
of $\partial {\mathcal C}_N/\partial \gamma^T$ as a function of $N\log(N)$, with $N$ being the system size, close to criticality.
The chosen parameter values and the color coding in this figure is the same as that in 
fig.(\ref{figxymodel0}) and we have taken here $\gamma^T=0.001$. 
The filled points are the numerical values and the straight line fits confirm the scaling behaviour 
$\partial {\mathcal C}_N/\partial \gamma^T \sim N\log(N)$. This is the same scaling as
obtained for the derivative of the Berry phase near the anisotropy transition line of the transverse XY model in \cite{Zhu}. 

An entirely similar analysis follows for the derivative $\partial {\mathcal C}_N/\partial h^T$ across the Ising transition
lines $h^T = \pm 1$,  where the gap in the 
energy spectrum comes from $k=0, \pi$, and at these lines, the derivative 
$\partial\Delta\theta_k/\partial h^T \sim 1/(\gamma^T\sin k)$, i.e., diverges as $1/k$ (or as $1/(\pi - k)$)
near $k=0$ (or $k=\pi$). Hence, when $\Delta\theta_k$ is finite, which can be checked to be always the case
when $h^R \neq \pm 1$, the derivative of the NC diverges at the Ising phase 
transition. There is, however, an interesting counter-example, as seen from the 
dashed blue line in the inset of fig.(\ref{figxymodel})(b). 

Recall that this corresponded to the initial state on
the Ising critical line, and we had used here $(\gamma^R=1/3, h^R = 1,\gamma^T= 1/2)$. 
Now, it is not difficult to check that if $h^R = h^T=1$, then we have for small $k$,
$\Delta\theta_k = k/4(1/\gamma^T - 1/\gamma^R) + {\mathcal O}(k^3)$. In that case, 
for finite non-zero values of $\gamma^R$ and $\gamma^T$, $\partial {\mathcal C}_N/\partial h^T$
{\it does not} diverge as the factor of $k$ is cancelled, but shows a finite discontinuity at the Ising transition, 
and the latter behaviour can be
traced to fact that for $h^R=1$, $\Delta\theta_k$ changes sign near $h^T=1$. This explains
the behaviour of the graphs in the inset of fig.(\ref{figxymodel})(b).

We note that when our reference state is not on the critical line $h^T=\pm 1$, 
$\partial {\mathcal C}_N/\partial h^T$ captures the information of the QPT
across the Ising critical lines, while it is known that the Ricci scalar curvature that follows from the QIM 
(presented for this model in eq.(\ref{linexy})) can be transformed to a form that is independent of $h$ (see
discussion after eq.(\ref{spacexy})) and thus does not show any special behavior at 
these lines \cite{Zanardi},\cite{TapoDickeModel}.
This is a distinguishing feature of the NC, and its derivative is a better indicator of the phase
transition compared to the Ricci scalar at the Ising transition in the transverse XY model. 

It remains to compute the scaling behaviour of $\partial {\mathcal C}_N/\partial h^T$. From our
previous discussion, it is gleaned that near $h^T=1$, the derivative 
$\partial {\mathcal C}_N/\partial h^T$ should again scale like $N\log N$. To check this 
explicitly, we have, in fig.(\ref{scaling})(b), plotted the derivative calculated numerically as a function
of $N\log(N)$ with $N$ being the system size as before. In these plots, the filled points are
the numerically obtained values, and the same parameter values and color coding is used, 
as in fig.(\ref{figxymodel}), and we have taken $h^T=1$. The linear nature of the fit confirms the scaling behaviour mentioned. 
\begin{figure}[h!]
\centering
\begin{subfigure}{0.48\columnwidth}
\includegraphics[width=\textwidth]{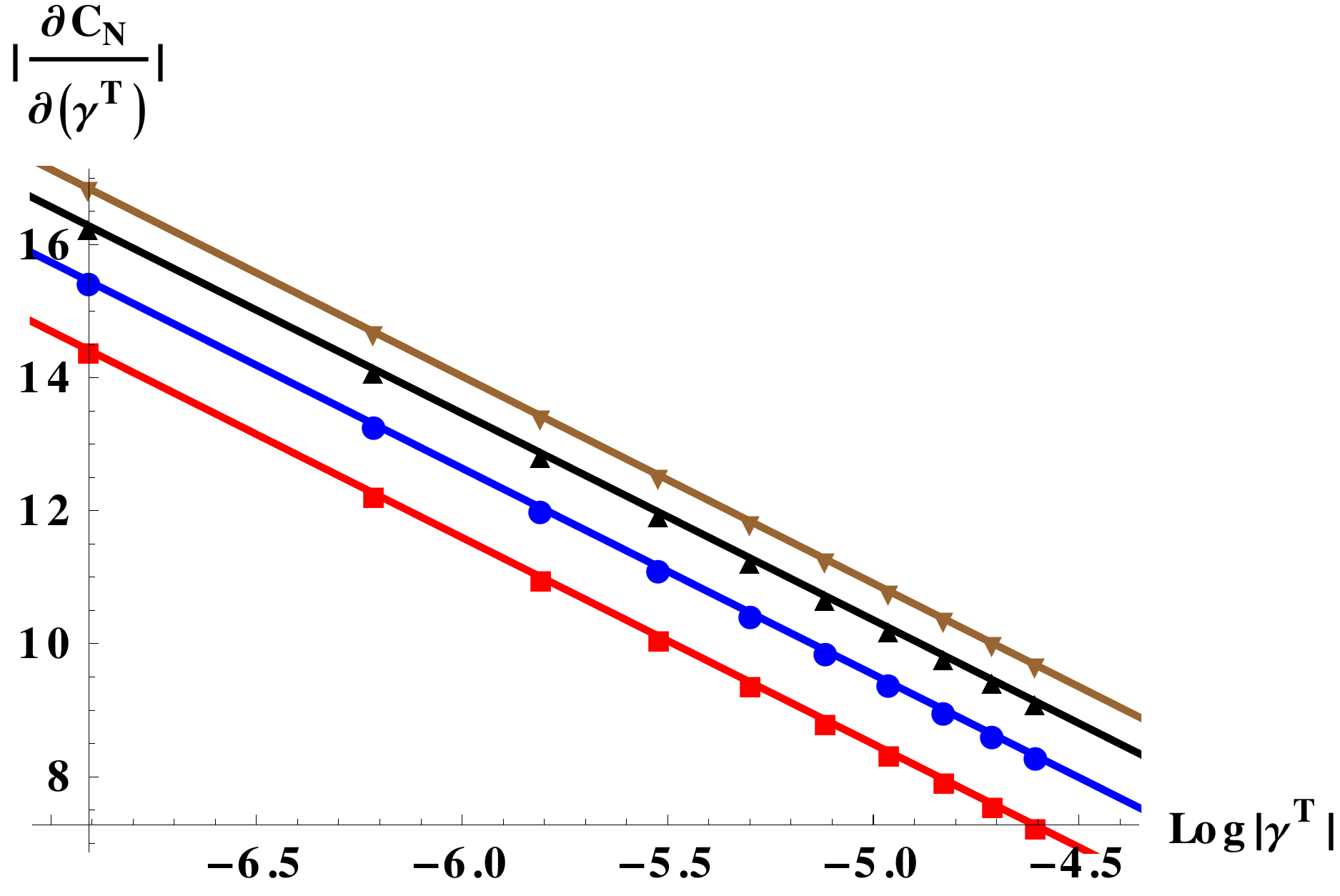}
\caption{Linear scaling of $\partial {\mathcal C}_{N}/\partial \gamma^T$ as a function of $\log|\gamma^T|$
near $\gamma^T=0$.}
\end{subfigure}
\hfill
\begin{subfigure}{0.48\columnwidth}
\includegraphics[width=\textwidth]{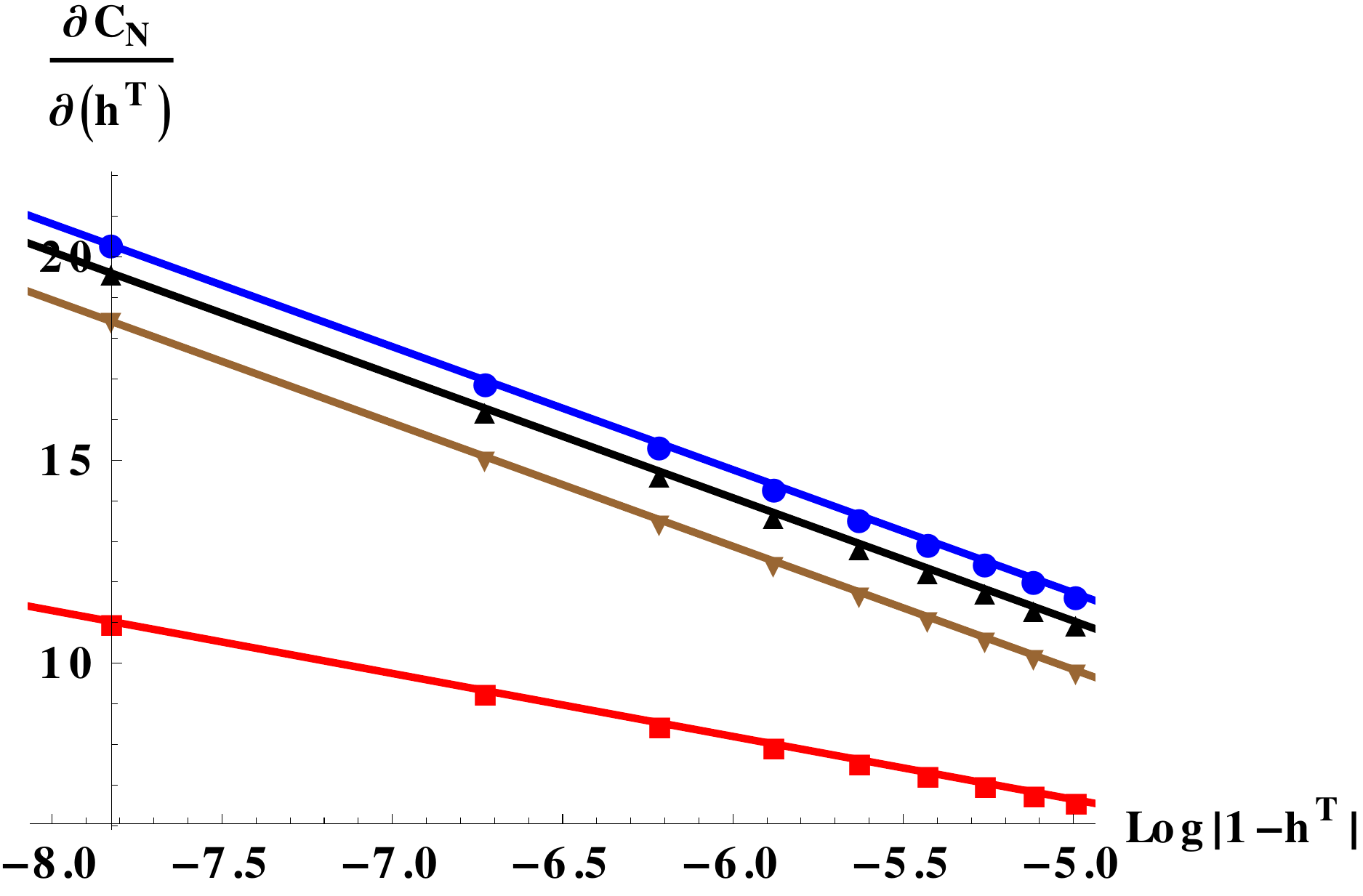}
\caption{Linear scaling of $\partial {\mathcal C}_{N}/\partial h^T$ as a function of $\log|1-h^T|$ near $h^T=1$.} 
\end{subfigure} 
\caption{Scaling of $\partial {\mathcal C}_{N}/\partial \gamma^T$ (a) and 
$\partial {\mathcal C}_{N}/\partial h^T$ (b) in the thermodynamic limit.} 
\label{scalingtd}
\end{figure}

Finally, it is important to establish the scaling of the derivative of the NC near the critical lines, as
a function of the system parameters, in the thermodynamic limit. 
We find that near $\gamma^T = 0$, $\partial {\mathcal C}_{N}/\partial \gamma^T \sim \log|\gamma^T|$ and that 
close to $h^T=1$, $\partial {\mathcal C}_{N}/\partial h^T \sim \log|1-h^T|$. These
are depicted in figs.(\ref{scalingtd})(a) and (b) (with the same parameter values and color codings as
in figs.(\ref{figxymodel0}) and (\ref{figxymodel}) respectively), where we have plotted these derivatives as functions of 
$\log|\gamma^T|$ and $\log|1-h^T|$ respectively. The filled points corresponding to numerical values and 
the solid lines giving the fits. The linear nature of the derivatives are visible. 
Again, these scaling behaviours are identical to those of the derivative of the Berry phase, obtained in \cite{Zhu}.

This logarithmic divergence of the derivative of the NC in fact has a nice mathematical 
explanation. It will be convenient to work with a single varying parameter, $\lambda^T$, which we will 
call $\lambda$ to prevent cluttering notation. 
From the expression in eq.(\ref{NC}), we obtain (in the same manner that we derived eq.(\ref{NCdiv})), 
$\partial {\mathcal C}_N/\partial \lambda^T \equiv \partial {\mathcal C}_N/\partial \lambda= 
\sum_k 2\Delta\theta_k(\partial\theta_k^T/\partial\lambda)$. Differentiating
once more, we get $\partial^2 {\mathcal C}_N/\partial \lambda^{2} = \sum_k 2(\partial\theta_k^T/\partial\lambda)^2 +
\sum_k 2\Delta\theta_k(\partial^2\theta_k^T/\partial \lambda^{2})$. Now, it can be checked by explicitly evaluating
the sums with $\lambda = \gamma^T$ or $h^T$, that near criticality, the two terms
appearing in the second derivative of the NC are of the same order, and that their ratio reaches a constant
value as one approaches a critical point (we will momentarily substantiate this). To obtain the scaling behaviour near criticality, 
it is then convenient to work only with the first term, which (up to an unimportant factor) 
is recognized to be the metric, from eq.(\ref{gij}). Then, we obtain near criticality, 
\begin{equation}
\frac{\partial {\mathcal C}_N}{\partial \lambda} \sim \int\sum_k\left(\frac{\partial\theta_k^T}{\partial\lambda}\right)^2d\lambda
= \int g_{\lambda\lambda}d\lambda~.
\label{ncder}
\end{equation}
In quasi-free fermionic systems, the most
relevant operator has a dimension of one. As shown in \cite{Venuti}, in these systems, in the thermodynamic
limit, close to criticality, this leads to $g_{\lambda\lambda} \sim |\lambda - \lambda_c|^{-1}$, where $\lambda_c$ is the
critical value of $\lambda$. Putting this in eq.(\ref{ncder}),
we obtain finally, 
\begin{equation}
\frac{\partial {\mathcal C}_N}{\partial\lambda} \sim \log|\lambda - \lambda_c|~,
\label{ncder1}
\end{equation}
confirming the validity of our numerical analysis presented in fig.(\ref{scalingtd}). 

\begin{figure}[h!]
\centering
\includegraphics[width=0.27\textwidth]{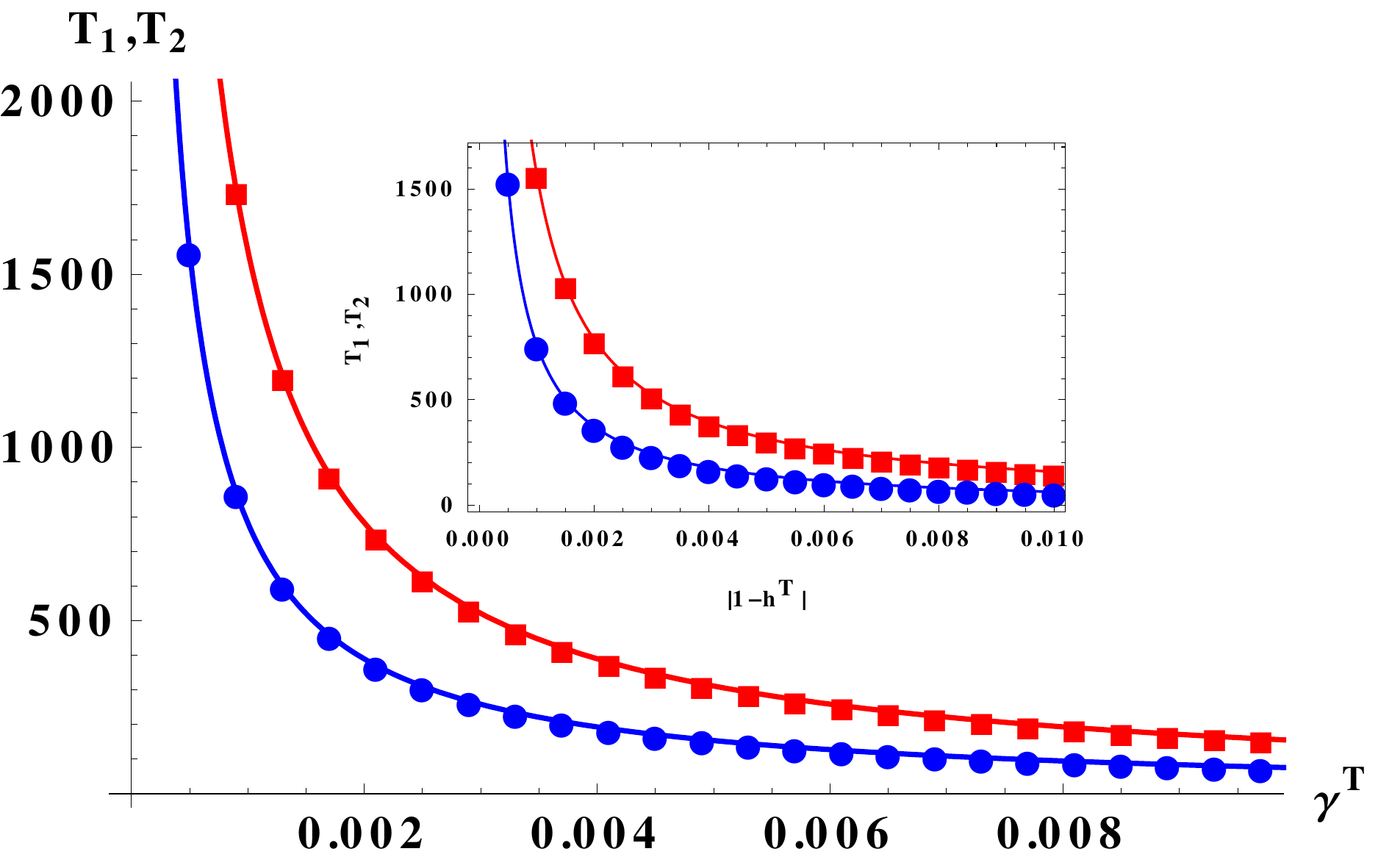}
\caption{Scaling of $T_1 = \sum_k (\partial\theta_k^T/\partial\lambda)^2$ (red) and 
$T_2 = \sum_k \Delta\theta_k(\partial^2\theta_k^T/\partial \lambda^{2})$ (blue)
with $\lambda = \gamma^T$ near $\gamma^T=0$. 
The inset shows the corresponding scaling for $\lambda = h^T$ near $h^T=1$.}
\label{scaling2}
\end{figure}

To substantiate our mathematical argument above, we should check that both terms in the second
derivative of the NC scale as $|\lambda - \lambda_c|^{-1}$ in the thermodynamic limit, as follows by taking a 
derivative of eq.(\ref{ncder1}). We call $T_1 = \sum_k (\partial\theta_k^T/\partial\lambda)^2$
and $T_2 = \sum_k \Delta\theta_k(\partial^2\theta_k^T/\partial \lambda^2)$ and consider these in the
thermodynamic limit. The scaling
of these two terms with $\lambda = \gamma^T$ near $\gamma^T=0$ is shown in the main fig.(\ref{scaling2}),
with $T_1$ in red and $T_2$ in blue. The filled points are the numerical values, and the solid lines are 
the corresponding fits with $T_1 = 1.57/\gamma^T + {\rm constant}$ and $T_2 = 0.79/\gamma^T + 
{\rm constant}$. The inset shows the corresponding fits for $T_1$ (red) and $T_2$ (blue) with 
$\lambda = h^T$, near $h^T=1$. In the inset, $T_1$ is fitted with the curve $1.57/|1-h^T| + {\rm constant}$
and $T_2$ with $0.77/|1-h^T| +  {\rm constant}$. This behaviour is of course independent of the initial
values chosen, and we have taken typical values $h^R=1/3$, $h^T=1/2$ and $\gamma^R=1/3$ in
the main figure (\ref{scaling2}), and $h^R=1/3$, $\gamma^R=1/3$ and $\gamma^T=1/2$ for the inset. 

We now turn to the FSC. The information metric for the transverse XY model has been 
calculated in \cite{Zanardi},\cite{Polkovnikov} and in the thermodynamic limit, 
the line element, in the ferromagnetic phase $|h| <1$, $\gamma>0$ is given by
\begin{equation}
ds^2 = \frac{dh^2}{16\gamma\left(1-h^2\right)} + \frac{d\gamma^2}{16\gamma\left(1 + \gamma\right)^2}~,
\label{linexy}
\end{equation}
with the replacement $\gamma \to -\gamma$ for the ferromagnetic phase with $\gamma < 0$. 
The geodesic equations for this model are given by \cite{TapoGeodesics}
\begin{equation}
{\ddot h} + \frac{h{\dot h}^2}{1-h^2} - \frac{{\dot h}{\dot \gamma}}{\gamma} = 0,~~{\ddot \gamma} - 
\frac{{\dot \gamma}^2\left(1+3\gamma\right)}{2\gamma\left(1 + \gamma\right)} + 
\frac{{\dot h}^2\left(1 + \gamma\right)^2}{2\gamma\left(1 - h^2\right)}=0~,
\label{geoxy}
\end{equation}
where the dot represents a derivative with respect to the affine parameter $\tau$. Also, the normalization 
condition implies that 
\begin{equation}
\frac{{\dot h}^2}{16\gamma\left(1 - h^2\right)} + \frac{{\dot \gamma}^2}{16\gamma\left(1 + \gamma\right)^2} = 1
\label{spacexy}
\end{equation}
In general, these might be complicated to solve, but an analytic solution is possible in the special case
$h={\rm constant}$. For this case, the first relation in eq.(\ref{geoxy}) is satisfied, and hence constant $h$ 
lines are geodesics. 
This observation is motivated by the fact that in eq.(\ref{geoxy}), $h$ is in fact a cyclic coordinate, as 
can be gleaned by a coordinate transformation $h = \sin\alpha$ for an angular variable $\alpha$ which 
is justified as we are in a region $|h| < 1$. Now
then, it is a simple matter to find $\gamma(\tau)$ from the second equation
in eq.(\ref{geoxy}), in conjunction with eq.(\ref{spacexy}), after putting ${\dot h}=0$ there. The former gives rise to
two undetermined constants, one of which is fixed from eq.(\ref{spacexy}). We finally find that 
$\gamma = \tan^2\left(2\left(\tau - \tau_0\right)\right)$, where $\tau_0$ is a constant reference value for the affine parameter.
An entirely similar result is obtained if we use the Euler-Lagrange equation corresponding to the cyclic 
coordinate $h$, algebraically solve it simultaneously with eq.(\ref{spacexy}), and solve the resulting
first order differential equations. 

Now, it can be seen that geodesics with $\tau_0=0$ reach the singularity at $\gamma = 0$, and using this value as
the origin of our geodesic, the final expression for the FSC is 
\begin{equation}
C_{FS}=\tau = (1/2)\arctan(\sqrt{|\gamma|})~, 
\end{equation}
for this class of geodesics.  
Note that the first derivative of the FSC diverges at the anisotropic phase transition $\gamma \to 0$ as
$d{\mathcal C}_{FS}/d\gamma \sim \gamma^{-1/2}$. This divergence of the 
first derivative of the FSC (much like the NC) signals a quantum phase transition. The power
law behaviour of the derivative of the FSC is to be contrasted with the logarithmic divergence
of the derivative of the NC that we have just seen. To see this, we have 
to only remember the expression of the line element of eq.(\ref{genline}), i.e.,
$d\tau^2 = g_{ij}d\lambda^id\lambda^j$, where $\tau$ is measured along a geodesic. 
Then, we have, for a given parameter $\lambda$,
\begin{equation}
\frac{\partial\tau}{\partial\lambda} = \frac{\partial{\mathcal C}_{FS}}{\partial\lambda} = 
\left(g_{\lambda\lambda}\right)^{\frac{1}{2}}\sim |\lambda - \lambda_c|^{-\frac{1}{2}}~,
\label{fscder}
\end{equation}
where, as before, $\lambda_c$ denotes the critical value of the parameter $\lambda$, 
and the last relation follows from \cite{Venuti}, as in our previous argument, given after eq.(\ref{ncder}). 
Note that eq.(\ref{fscder}) is a generic scaling behaviour, as it follows directly from the 
expression of length, and will also be valid for non-geodesic paths. 
We have considered here the region
$|h|<1$. The geometry of the paramagnetic phase with $|h|>1$ turns out to be way too complicated to
handle analytically. 

Of course our computation here was facilitated by the fact that we could guess a solution for one of the
variables, namely $h$. 
It is instructive to understand FSC from a perspective in which we could have alternatively achieved this. 
Remember that we had introduced a rotation of the model
about the $z$-axis via the transformation given in eq.(\ref{rotHamiltonian}). 
Then it can be checked that there is a non-zero metric component along the 
$\phi$ direction, which, from the expression for the ground state in eq.(\ref{gsXY}), 
can be straightforwardly shown to be
$g_{\phi\phi} = \sum_k\sin^2\theta_k$, upon using eq.(\ref{genline}). An explicit computation gives 
the following metric on the $\gamma-\phi$ plane for $|h| <1$,
\begin{equation}
ds^2 = \frac{|\gamma| d\phi^2}{2\left(1+|\gamma|\right)} + \frac{d\gamma^2}{16|\gamma|\left(1 + |\gamma|\right)^2}~,
\label{linexyphi}
\end{equation}
which is a space of constant scalar curvature for $|\gamma| >0$ and has a delta function singularity at $\gamma = 0$. 
This is a particularly convenient form of the metric as the coordinate $\phi$ is cyclic, and we will focus on 
$\gamma >0$. In this plane, we proceed by noting that as mentioned before, 
the geodesic equations are given via the Euler-Lagrange equations
for a Lagrangian ${\mathcal L} = 1/2\sqrt{(g_{ij}{\dot\lambda^i} {\dot\lambda^j})}$ with now ${\vec\lambda} = (\gamma,\phi)$. 
However, since $\phi$ is cyclic, we
have $\partial{\mathcal L}/\partial{\dot\phi}={\rm constant}=K$. This equation, along with the 
normalization condition for the vector $u^i = ({\dot\phi},{\dot\gamma})$, namely $g_{ij}u^iu^j=1$ 
(see discussion after eq.(\ref{Lagrangian})), can be algebraically solved to obtain $u^i$. The solution is 
\begin{equation}
{\dot\phi}=\frac{2K\left(1+\gamma\right)}{\gamma}~,~~{\dot\gamma}=-4\left(1+\gamma\right)\sqrt{\gamma-2K^2(1+\gamma)}
\label{phidot}
\end{equation}
At this stage, since the complexity does not have any contribution from $\phi$, 
we set $K=0$ so that we consider paths with constant $\phi$, which are geodesics as can be 
checked from eq.(\ref{linexyphi}). Setting this constant to zero also ensures that the affine 
parameter can be set to zero at the phase transition. 
Then, the second relation in eq.(\ref{phidot}) can be solved for $\gamma$ as a function of the
affine parameter.  
There will be one undetermined constant here, which can be fixed by demanding that
the geodesics reach the critical line $\gamma = 0$ at $\tau = 0$.

Performing this exercise, we finally obtain $|\gamma| = \tan^2\left(2\tau\right)$, as before. 
The FSC computed in the $\gamma-\phi$ plane again equals $(1/2)\arctan(\sqrt{|\gamma|})$, and will have the
same behaviour as what we had obtained in the $h-\gamma$ plane with $|h|<1$. 
The reason why the FSC matches in these cases is
that the geodesic equation here is formally the same as the second one in eq.(\ref{geoxy}) if we choose
$\phi$ to be a constant. Alternatively, 
we can make a change of variables $\gamma = \tan^2\alpha$ (for $\gamma >0$) and note that
the line element is that of a sphere with a conical defect at $\alpha=0$. Now geodesics being
great circles, the geodesic length is proportional to $\alpha$, which gives the desired result. 
The analysis in the $\gamma-\phi$ plane for $|h|>1$ is substantially more complicated and we will
not comment on this. 

Finally, we focus on the $h-\phi$ plane. From the metric components $g_{hh}$ and $g_{\phi\phi}$
given in eqs.(\ref{linexy}) and (\ref{linexyphi}), it is 
checked that in the region $|h|<1$, the metric is flat, i.e. geodesics are straight lines. The flatness of
the metric is seen by the coordinate transformation $h=\sin\alpha$, as mentioned before. 
The geodesic length here will be proportional to $\alpha$, which in terms of the original 
coordinates gives an inverse sine function whose derivative will diverge in the limit $h\to1$ as
$\sim |1-h|^{-1/2}$, as expected. 
In the region $|h|>1$, it is difficult to find an exact analytic solution for the geodesic length, as the 
expressions for the metric components become cumbersome. However, in the
limit that $h$ is close to unity, we find, after a similar analysis as above 
that the geodesic length, i.e., the FSC for non-zero $\gamma$ is 
\begin{equation}
{\mathcal C}_{FS} = \frac{\sqrt{|h|-1}}{2\sqrt{2|\gamma|}}~.
\end{equation}
This equation is derived by taking the metric components $g_{hh}$ and $g_{\phi\phi}$ for 
$|h|>1$ as derived in \cite{Polkovnikov}, expanding these around $h=1$ and then computing
the geodesic equation with the approximated metric. 
Near $|h| \to 1$, the derivative of the FSC diverges as $(|h|-1)^{-1/2}$, in lines with
eq.(\ref{fscder}).  

Above, we have analysed the derivative of the FSC for some special geodesics. Clearly, it is of importance
to understand generic ones, that satisfy eq.(\ref{geoxy}). Analytical formulas are difficult to obtain
in these cases, and we will resort to a numerical study. We parametrise $h$ and $\gamma$ by the affine
parameter $\tau$, and solve the two second order equations of eq.(\ref{geoxy}) by specifying four
initial conditions. As a concrete example, we will choose here the initial conditions 
$h=0.16$, $\gamma=0.3$, ${\dot \gamma} = -0.1$. Then, ${\dot h}$ is determined from the normalization
condition of the tangent vectors ${\dot h}$ and ${\dot \gamma}$ (see discussion after eq.(\ref{Lagrangian})),
and is given by ${\dot h} = 2.16$ here. Note that the normalization condition has to be satisfied for
all values of the affine parameter. We have checked that this is true for the numerical solutions that
we obtain. 

\begin{figure}[h!]
\centering
\begin{subfigure}{0.48\columnwidth}
\includegraphics[width=\textwidth]{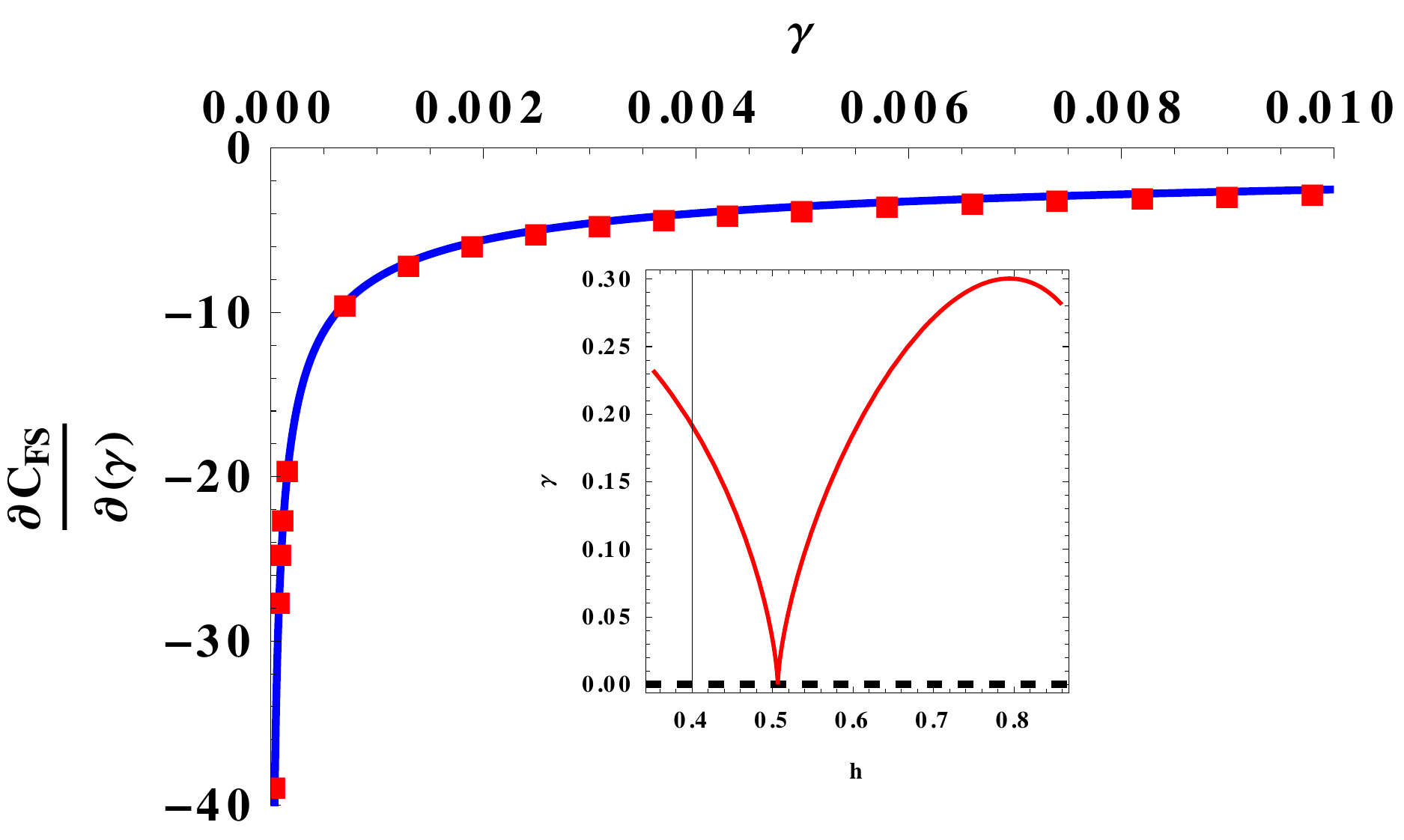}
\caption{$\partial {\mathcal C}_{FS}/\partial \gamma$ near $\gamma=0$ in the XY model. The inset shows a geodesic
near $\gamma=0$} 
\end{subfigure} 
\hfill
\begin{subfigure}{0.48\columnwidth}
\includegraphics[width=\textwidth]{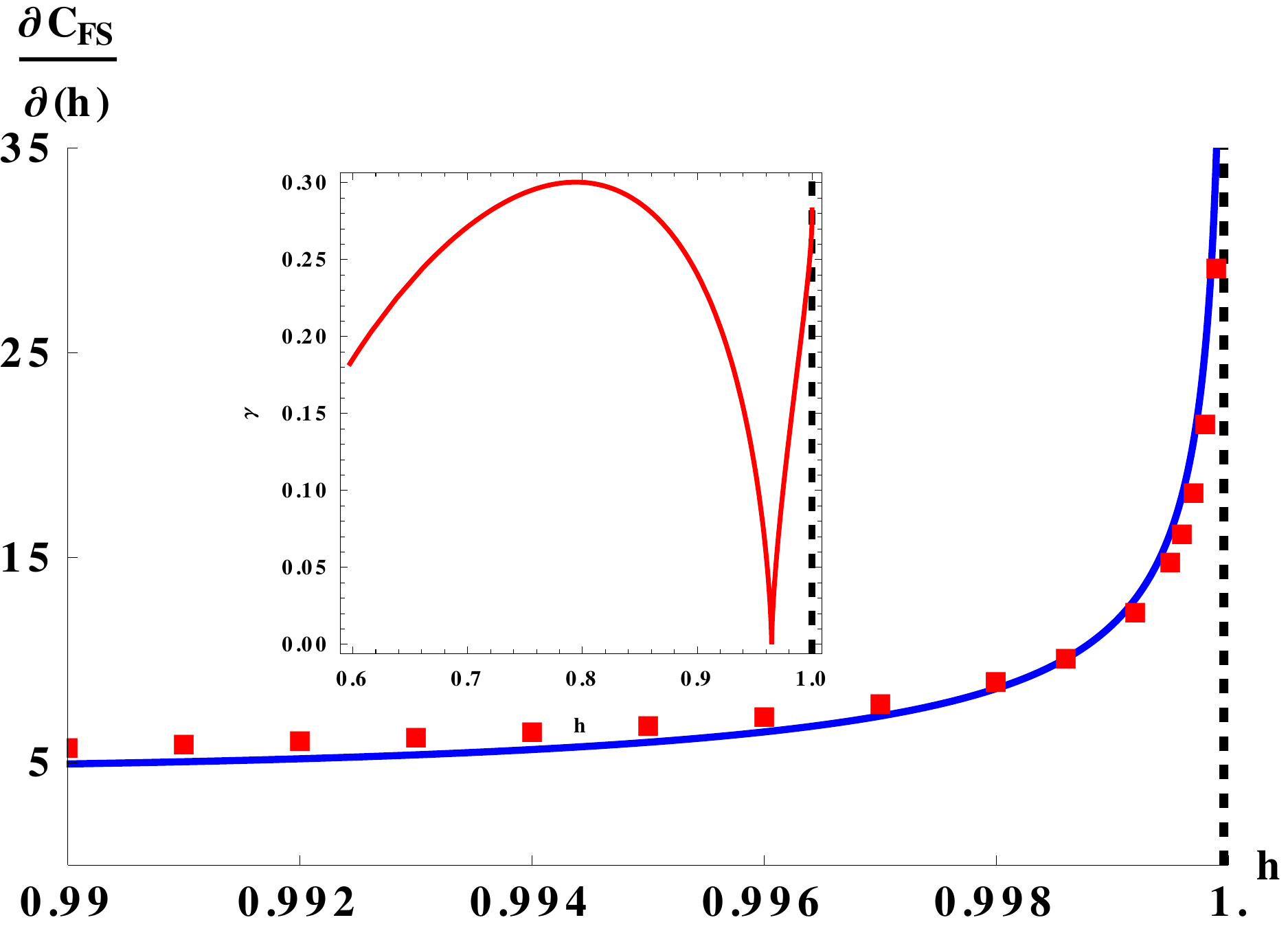}
\caption{$\partial {\mathcal C}_{FS}/\partial h$ near $h=1$ (dotted black line) in the XY model. The inset shows the geodesic
near $h=1$}
\end{subfigure}
\caption{The derivative $\partial {\mathcal C}_{FS}/\partial \gamma$ (a) and 
$\partial {\mathcal C}_{FS}/\partial h$ (b). The insets show a geodesics on the $h-\gamma$
plane near the phase boundaries (see text).} 
\label{gengeo}
\end{figure}

Once we find the numerical solution of the geodesic equations,
i.e., obtain $h(\tau)$ and $\gamma(\tau)$ as independent interpolating functions, we can invert them by 
a standard root finding procedure in Mathematica, which yields the affine parameter $\tau(h)$ and $\tau(\gamma)$, 
which is the FSC (as a function of $h$ or of $\gamma$) from our previous discussion, and here $\tau$ 
is measured from the starting point 
of the geodesic that is determined from the initial conditions. Now, one can numerically obtain the derivatives 
of the FSC. We show this in figs.(\ref{gengeo})(a) and (b), where the 
derivative of the FSC is plotted with the solid blue lines in the figures. The insets of
these figures show the geodesic plotted on the $h-\gamma$ plane (these have been discussed in 
\cite{TapoGeodesics}) near $\gamma = 0$ in fig.(\ref{gengeo})(a) and near $h=1$ in fig.(\ref{gengeo})(b). 
As can be clearly seen from the figures, the derivative of ${\mathcal C}_{FS}$
diverges near the critical lines $\gamma = 0$, and $h=1$. The red filled points in figs.(\ref{gengeo})(a) and (b)
correspond to the fits $\partial{\mathcal C}_{FS}/\partial\gamma=-0.24|\gamma|^{-1/2} + {\rm constant}$ and
$\partial{\mathcal C}_{FS}/\partial h=0.26|1-h|^{-1/2}+ {\rm constant}$, and confirm
eq.(\ref{fscder}).

\section{\label{compass} The quantum compass Model}

Now we study complexity in the one-dimensional quantum compass model \cite{Perk}, \cite{Compass0}, \cite{Compass1}, 
\cite{Compass2}, \cite{Jaffari1}, \cite{Jaffari2}. The two-dimensional
lattice version of this model has been studied for a long time, while its one-dimensional {\it avatar} is relatively 
recent, and was introduced in \cite{Compass0} to study an analytic model of a quantum magnet that showed
a first order phase transition. This was achieved by a diagonalization procedure in a subspace of the 
full Hilbert space in which the Hamiltonian reduces to that of the quantum Ising model. In \cite{Compass2},
a modified form of the one-dimensional compass model was considered, and it was found that with two
tunable parameters, the model showed a second order phase transition line, along with the first order one. 
This model is analogous to a one dimensional XY spin with alternating interactions, and is the natural
extension of the model considered in the previous section. 

Here, the Hamiltonian for $N=2N'$ spins is defined by 
\begin{equation}
{\cal H } = \sum\limits_{n=1}^{N'}\left[J_1\sigma^z_{2n-1}\sigma^z_{2n} + J_2\sigma^x_{2n-1}\sigma^x_{2n} 
+ L_1\sigma^z_{2n}\sigma^z_{2n+1}\right]~,
\label{compassnoH}
\end{equation}
and we closely follow the notation of \cite{Compass2}. There, the ground state energy in the thermodynamic limit was computed, 
and the existence of a second order phase transition line (along with a first order one) was explicitly demonstrated
from the (dis)continuity of the first derivative of the ground state energy. 

Again, for our purposes, it is useful to introduce a rotation $\phi$ about the $z$-axis, 
defined by eq.(\ref{rotHamiltonian}). By a standard diagonalization procedure, the ground state is given as 
\begin{equation}
\ket{g} = \prod\limits_{k=0}^{\pi}\cos\left(\frac{\theta_k}{2}\right)\ket{0}_k\ket{0}_{-k} 
-i\sin\left(\frac{\theta_k}{2}\right)e^{-2i\phi}\ket{1}_k\ket{1}_{-k}~,
\label{gsCompass}
\end{equation}
where, as before, $\ket{0}_k$ and $\ket{1}_k$ are the vacuum and single excitation states of Jordan-Wigner
fermions with momentum $k$. We find that the Bogoliubov angle $\theta_k$ is given here as
\begin{equation}
\cos\theta_k = \frac{\left(\frac{J_2}{L_1} - \cos k\right)}{\sqrt{\left(\frac{J_2}{L_1} - \cos k\right)^2 + \sin^2 k}}
\end{equation}
Note that $J_1$ does not enter into the expression for the ground state, and the parameter manifold
consists of the coordinates $J_2/L_1$ and $\phi$. The energy gap is 
\begin{equation}
\Delta_k = 4L_1\sqrt{1+(J_2/L_1)^2 - 2(J_2/L_1)\cos k}
\end{equation}
which closes at $k=0,\pi$ for $J_2/L_1=\pm 1$, indicating the location of QPTs. 
The NC of the ground state can be computed in the same way as in the last section, given
the Bogoliubov angle $\theta_k$. Note that $\theta_k$ here is {\it formally} identical with the Bogoliubov angle that appears
in the transverse XY model of eq.(\ref{Bogoliubov}) if we identify $h\to j$ and $\gamma \to 1$, 
although the models are distinct in this limit. 

Since the NC and its derivatives depend only on $|\Delta\theta_k|$, it is clear that 
for the compass model, the behaviour of the NC and its derivative will be identical to those
shown by the solid red curve in figs.(\ref{figxymodel})(a) and (b) (the transverse XY model 
with $\gamma = 1$). It is readily seen that similar to the transverse XY model, the scaling of the
$\partial {\mathcal C}_{N}/\partial j^T$ will be $\sim N\log(N)$ with $N$ being the system size, 
and in the thermodynamic limit, this derivative will scale as $\log|1-j^T|$. These should be 
clear from the behaviour of the red lines in figs.(\ref{scaling})(b) and (\ref{scalingtd})(b), respectively,
(with an appropriate change of notation mentioned above).

\subsection{Information metric of the ground state}

In our model of eq.(\ref{compassnoH}), taking the tunable parameters as \(j=J_2/L_1\) and \(\phi\), the
QIM takes the form
\begin{equation}
g_{jj}=\frac{1}{4}\!\sum\limits_{k=-\pi}^{\pi}\!\left(\frac{\partial\theta_{k}}{\partial j}\right)^{2},~~
g_{\phi\phi}=\sum\limits_{k=-\pi}^{\pi}\!\sin^{2}(\theta_{k})~,
\end{equation}
where, as before, the second relation follows from eq.(\ref{genline}) upon using the form of the ground state. 
The procedure to evaluate the QIM is straightforward, and the final results are
\begin{eqnarray}
g_{jj} &=& \frac{1}{16\left(1-j^2\right)}~,~~g_{\phi\phi} = \frac{1}{4}~,~|j|<1\nonumber\\
g_{jj} &=& \frac{1}{16j^2\left(j^2-1\right)}~,~~g_{\phi\phi} = \frac{1}{4j^2}~,~|j|>1~.
\label{compassmet}
\end{eqnarray}
We should  mention here that $g_{\phi\phi}$ in eq.(\ref{compassmet}) cannot be obtained 
by setting $\gamma = 1$ in the corresponding result for the transverse XY model reported
in \cite{Polkovnikov} which diverges in this limit, and the contour integrals here have to be
performed after setting $\gamma = 1$.
For this diagonal form of the metric, one can compute the Ricci scalar curvature from eq.(\ref{scalarcurvature})
with $x^1=j$, $x^2=\phi$, and $g$ being the determinant of the metric. We find that while the 
scalar curvature vanishes for $|j|<1$ indicating that the metric is flat there, it is a constant ($R = 32$ in
appropriate units) for $|j| >1$. 
From the point of view of the scalar curvature $R$, its discontinuity at $j=\pm 1$ thus signals a 
quantum phase transition.

\subsection{Fubini-Study complexity of the ground state}

In this case, the Bogoliubov angle depends on a single parameter $j$. So, the complexity will be computed
by a geodesic in the $j-\phi$ plane with constant $\phi$, in lines on our previous argument that 
the $g_{\phi\phi}$ component of the metric does not contribute to the complexity. 
The geodesic length in the $j-\phi$ plane can be obtained straightforwardly from the metric of eq.(\ref{compassmet}). 
Since the region $|j|<1$ is flat, the geodesics are straight lines. This case is similar to the one we had
discussed in the context of the transverse XY model. The flatness of the metric can be gleaned 
from a coordinate transformation $j = \sin\alpha$, where $\alpha$ is an angular variable. Then, in terms
of the original variables, geodesic lengths are proportional to $\arcsin(j)$ and its derivative with respect
to $j$ will diverge at the critical lines $j=\pm 1$, as $\sim |1\mp j|^{-1/2}|$, as before. 

We will then focus on the outer region $|j|>1$. Here, we find that the geodesic equations are 
\begin{equation}
{\ddot \phi} - \frac{2{\dot j}{\dot\phi}}{j}=0~,~{\ddot j} +\frac{4\left(j^2 -1\right){\dot \phi}^2}{j}
- \frac{\left(2j^2-1\right){\dot j}^2}{j\left(j^2-1\right)}=0
\label{geocompass}
\end{equation}
Clearly, $\phi = {\rm constant}$ lines are geodesics and these do not contribute to the complexity
as required. In the same 
way as in the previous section, we can solve for $j$. Performing a similar computation as before,
we finally obtain here,
\begin{equation}
{\mathcal C}_{FS} = \frac{1}{4}\left(\frac{\pi}{2}-\arcsin\left(\frac{1}{j}\right)\right)~,~|j|>1~,
\label{cfs1}
\end{equation}
where we have taken the convention that the geodesic path originates from the critical point
at $j=1$, in the $j-\phi$ plane. The derivative of the FSC blows up when $j\to 1$, 
as $\sim |j-1|^{-1/2}$, as expected.
Eq.(\ref{cfs1}) can be understood as in the similar discussion in the XY model. Namely,
if we substitute $j=1/p$ in the region $|j|>1$ and follow this by a further coordinate
transformation $p=\sin\alpha$ for an angular variable $\alpha$, then with a redefinition of
$\phi$, the metric becomes that of the two sphere with polar angle $\alpha$. Then the 
length of the geodesics (great circles) are proportional to $\alpha$, which yields eq.(\ref{cfs1}). 
\begin{figure}[h!]
\centering
\begin{subfigure}{0.4\columnwidth}
\includegraphics[width=\textwidth]{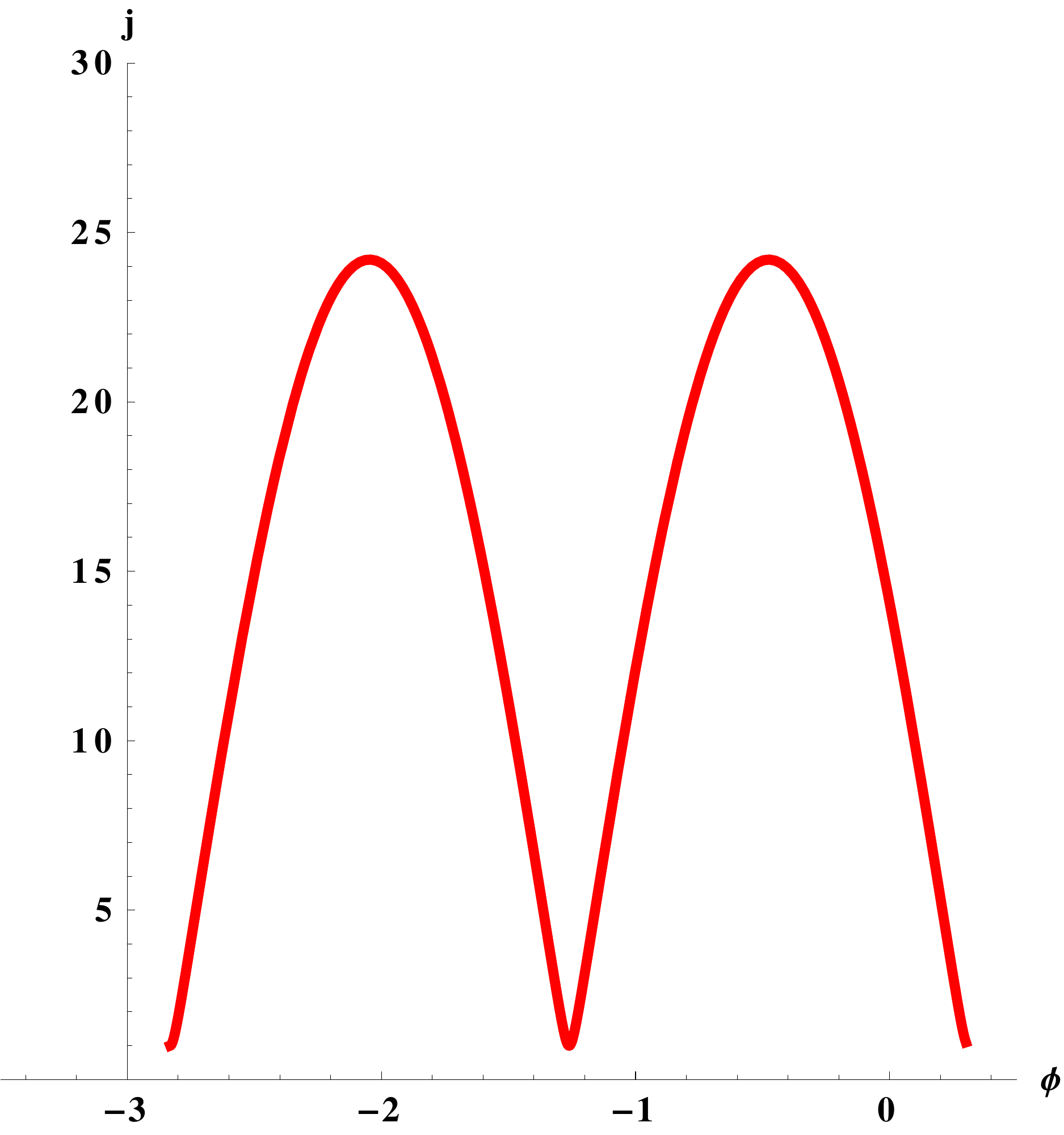}
\caption{A typical geodesic of the compass model in the $\phi - j$ plane.}
\end{subfigure}
\hfill
\begin{subfigure}{0.48\columnwidth}
\includegraphics[width=\textwidth]{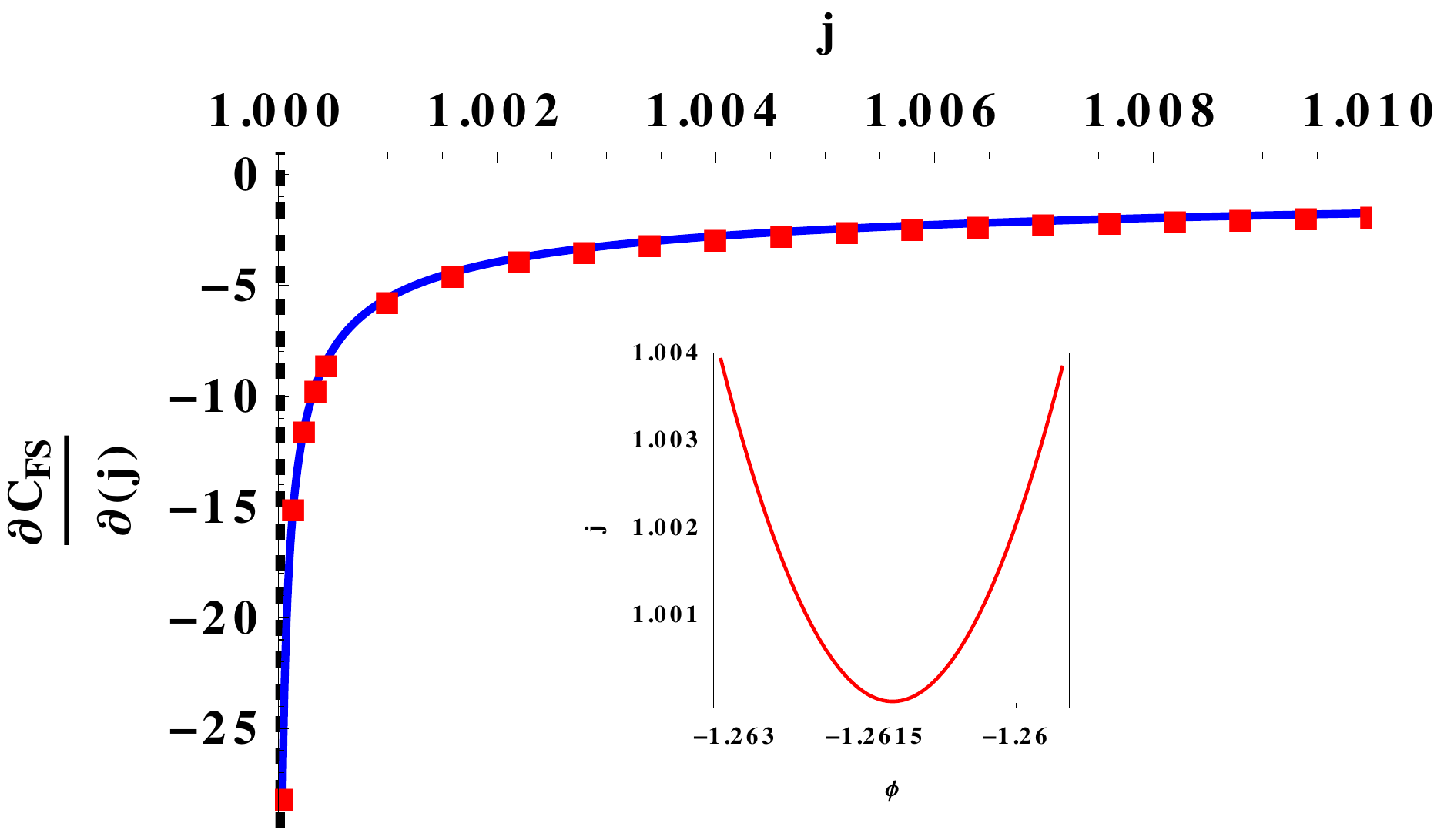}
\caption{Behaviour of $\partial{\mathcal C}_{FS}/\partial j$ in the compass model.} 
\end{subfigure} 
\caption{(a) A typical geodesic in the $\phi - j$ plane in the compass model. (b) The behaviour of 
$\partial{\mathcal C}_{FS}/\partial j$ in the compass model. The inset of (b) shows 
the geodesic near the turning point $j=1$ (see text).}
\label{fig2a}
\end{figure}  

In the above, we have taken $\phi = {\rm constant}$ lines as our geodesics. We can also analyze
the situation for more generic geodesics by solving eq.(\ref{geocompass}) numerically. This is
shown in fig.(\ref{fig2a})(a). Here, we have started with the initial conditions $j=1.1$, $\phi = 0.3$,
${\dot\phi}=-0.1$, with the normalization condition of the tangent vectors yielding ${\dot j} = 2.01$
as the fourth initial condition. The geodesic is plotted in the $\phi - j$
plane, where a periodic behaviour is seen. As has been observed in previous works \cite{TapoGeodesics}, 
the geodesic does not cross the phase boundary at $j=1$ and shows a turning behaviour near this. 

A standard numerical procedure is now used to 
invert the geodesic equations to obtain the affine parameter ${\mathcal C}_{FS}$, which is
measured from the starting point. Its derivative with respect to $j$ close to $j=1$ is shown in fig.(\ref{fig2a})(b),
where the divergence at $j=1$ (the dashed vertical line) is gleaned, with the filled
red points correspond to the fit $\partial{\mathcal C}_{FS}/\partial j = -0.18|j-1|^{-1/2} + {\rm constant}$, 
again confirming eq.(\ref{fscder}). The inset of fig.(\ref{fig2a})(b) shows a close up of 
the geodesic near $j=1$, $\phi \simeq -1.26$, where it shows turning behaviour. 

\section{\label{compassModH} Compass model in a transverse magnetic field}

We now consider complexity of the compass model in a transverse magnetic field, with periodic boundary
conditions. The spectrum of this model, in which we introduce cells with double sites (denoted by $1$ and $2$ in sequel), 
was studied in \cite{SunChen}, and \cite{Wang} analyzed the Berry phase of this model. Our purpose
here is to calculate the NC and the FSC for this model, for which we will need to compute the
information metric. We note that in this model, introduction of the transverse magnetic field does not
yield any additional phase structure. However, the NC and the FSC near criticality will non-trivially
depend on the magnetic field, as we will see. 

We start with the Hamiltonian for this model, which is given by the expression \cite{SunChen} 
\begin{equation}
H = -\sum\limits_{n=1}^{N'}\left[J_1\sigma^x_{2,n}\sigma^x_{1,n+1} + J_2\sigma^y_{1,n}\sigma^y_{2,n} 
+ \lambda\left(\sigma^z_{1,n}+ \sigma^z_{2,n}\right)\right]~,
\label{compassH}
\end{equation}
where $N = 2N'$ is the total number of sites, $\sigma_{1(2),n}$ are the Pauli matrices on cell
$n$ with sites $1(2)$, and $\lambda$ is a transverse magnetic field. We will 
work with two independent couplings $J_2/J_1$ and $\lambda/J_1$. We will 
finally scale the coupling $J_1=1$. We also rotate the Hamiltonian about the $z$-axis much in the same way
as in eq.(\ref{rotHamiltonian}). 

The derivation of the NC proceeds as before, and is provided in eq.(\ref{NCcompassH}) in appendix \ref{AppendixB}.
We simply mention that the NC
shows the typical behaviour that we have encountered before, namely its derivative diverges
at the location of the phase transitions, indicating its non-analyticity. Since this behaviour is identical to
the cases that we have already considered, 
we will not belabor upon the details further. We will rather look at the more interesting case of the FSC, via the information metric. 

\subsection{Information metric of the ground state}

We first discuss the information metric of the ground state. As mentioned before, without loss of
generality, we will scale $J_1=1$ and use the variables $j=J_2/J_1$ and $h=\lambda/J_1$. The components
of the QIM yield analytic expressions and we present the results for $g_{hh}$ and $g_{\phi\phi}$. In the
thermodynamic limit, these are
\begin{eqnarray}
g_{hh} &=& \frac{\left(1-j^2\right)^2 + 4h^2\left(1+j^2\right)}{4\left((1+4h^2+j^2)^2 - 4j^2\right)^{3/2}}~,\nonumber\\
g_{\phi\phi} &=& \frac{\sqrt{\left(1+4h^2+j^2)^2 - 4j^2\right)}- 2h^2}{4\sqrt{(1+4h^2+j^2)^2 - 4j^2}}~,
\label{metcomph}
\end{eqnarray}
and valid for all values of $j$. The quantity $g_{jj}$ is more cumbersome and have different values in the
regions $|j|<1$ and $|j|>1$ and these will not be presented here. 
  
\begin{figure}
\centering
\begin{subfigure}{0.48\columnwidth}
\includegraphics[width=\textwidth]{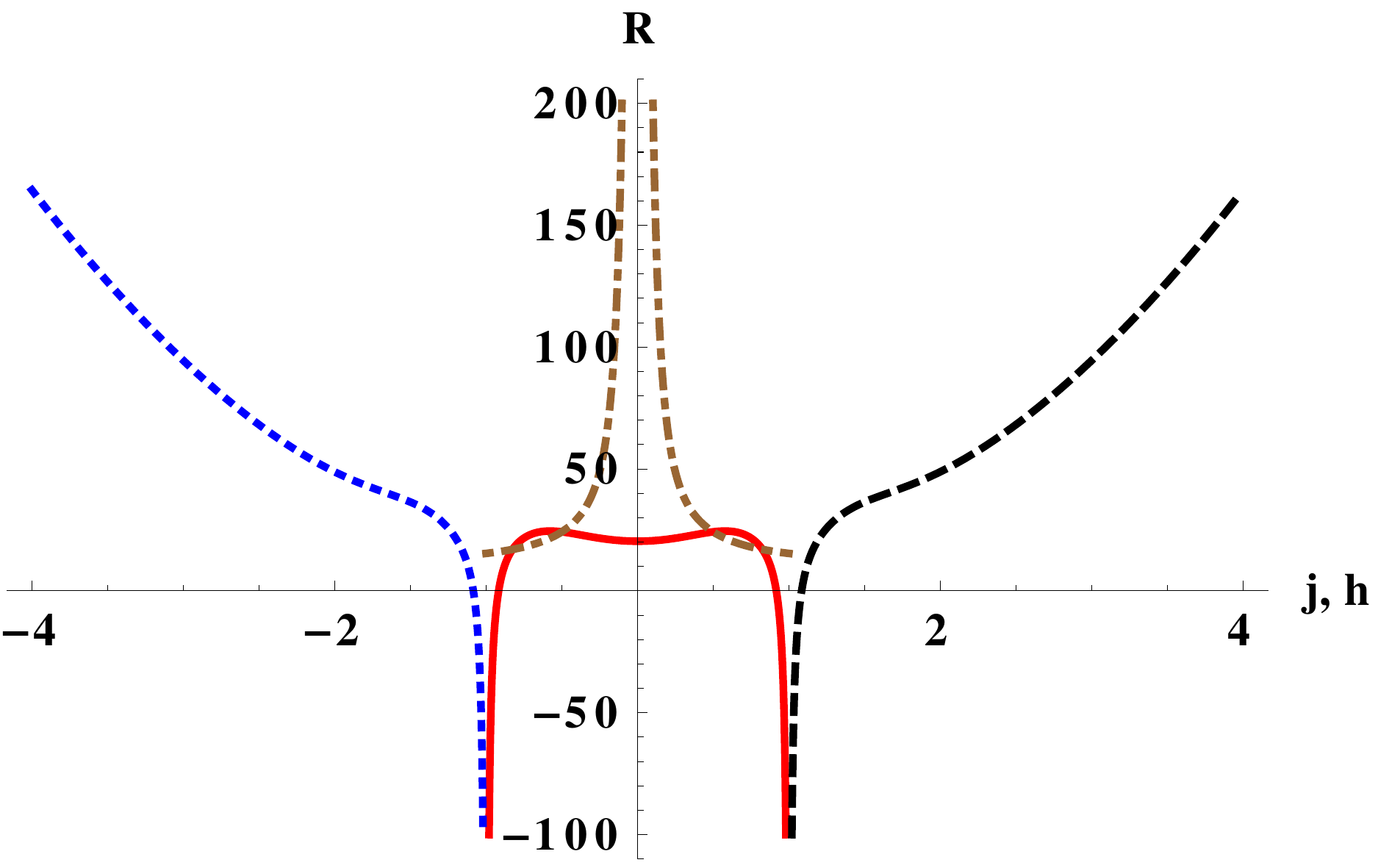}
\caption{$R$ vs $j$, in the $h-j$ plane, with $h=1/2$ in the ranges $|j|<1$ (red), $j<-1$ (dotted blue), 
$j>1$ (dashed black) and $R$ vs $h$, $j=1/2$, (dot-dashed brown).}
\end{subfigure}
\hfill
\begin{subfigure}{0.48\columnwidth}
\includegraphics[width=\textwidth]{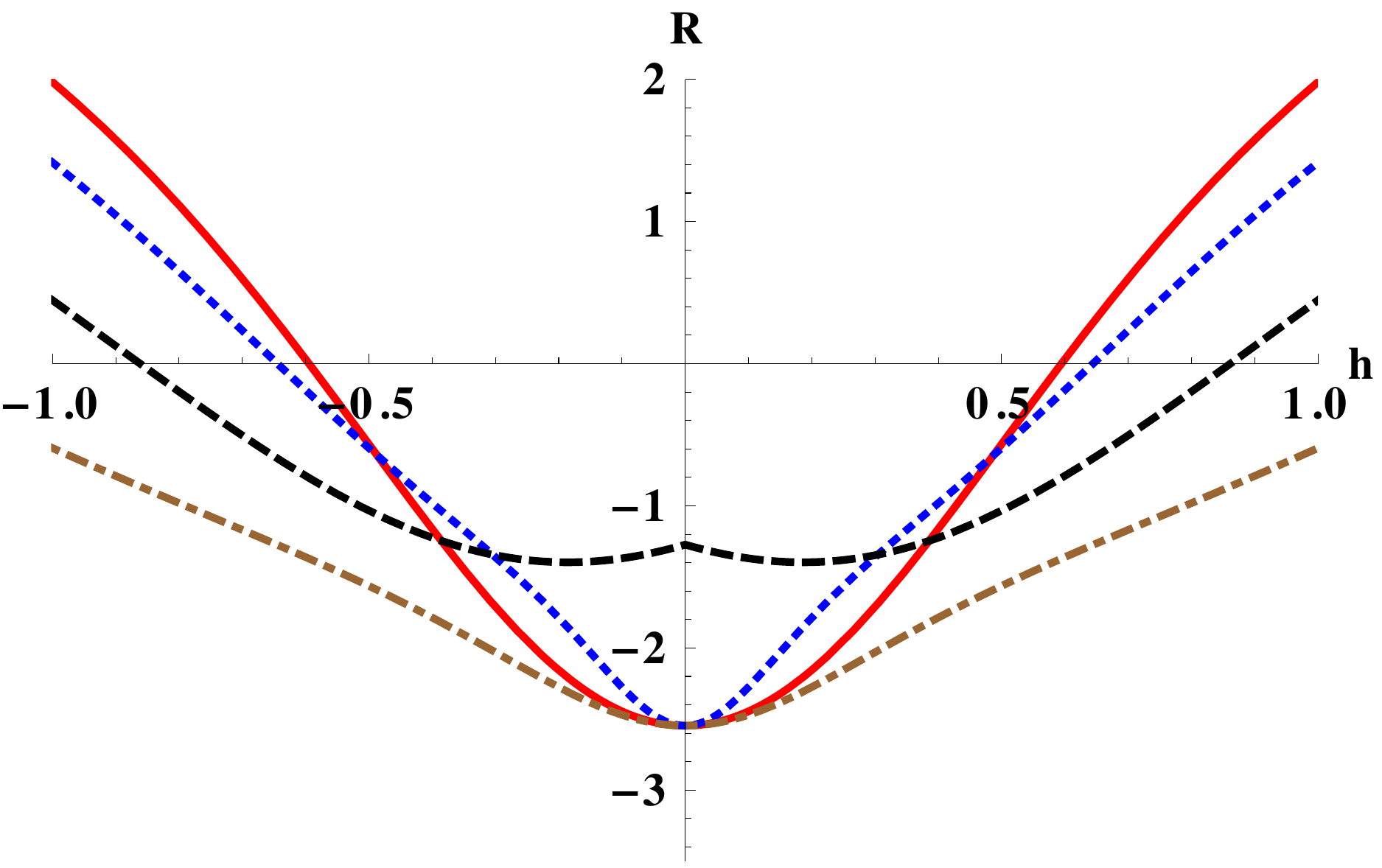}
\caption{$R$ vs $h$ in the $h-\phi$ plane, with $j=0$ (red), $1/2$ (dotted blue), $1$ (dashed black) 
and $2$, (dot-dashed brown) } 
\end{subfigure} 
\caption{Behaviour of the scalar curvature for the compass model in a transverse magnetic field, given by the
Hamiltonian of eq.(\ref{compassH}).} 
\label{fig3}
\end{figure}

First, we will analyze the scalar curvature $R$ arising out of the metric graphically, since the
analytic expression turns out to be lengthy. In fig.(\ref{fig3})(a), 
we have shown $R$ in the $h-j$ plane as a function of $j$, and as a function of $h$, for values indicated in
the figure. While $R$ diverges at the second order phase transition $j = \pm1$, it shows a spurious 
divergence at $h=0$ for any value of $j$. Such divergences associated with $R$ are known 
in the literature, and here it happens due to the fact that at $h=\lambda=0$, the energy gap 
closes partially as is evident from eq.(\ref{engap}). Fig.(\ref{fig3})(b) shows the scalar curvature in the
$h-\phi$ plane, and indicates regularity everywhere. 

\subsection{Fubini-Study complexity of the ground state}

We now study the FSC of the ground state. Even though the coordinate $\phi$ is cyclic, and $\phi={\rm constant}$
lines are geodesics, 
it is difficult to obtain an exact analytic formula for the geodesic length in the $h-\phi$ plane 
in this case, given the complicated nature of the metric components
of eq.(\ref{metcomph}). However, we find that this can be done
for small values of $h$, following the procedure discussed earlier. Namely, we can obtain an exact algebraic expression
for ${\dot h}$ from the normalization condition and the fact that $\phi$ is cyclic, and the resulting first order differential equation can 
be analytically solved for small $h$. We will skip all cumbersome details, and simply
mention that the solution for the geodesic length, i.e. the FSC turns out to be 
\begin{equation}
{\mathcal C}_{FS} = \frac{\sqrt{1-j^2}}{4\sqrt{1+j^2}}\arctan\left(\frac{2h\sqrt{1+j^2}}{1-j^2}\right)~,
\end{equation}
for small values of $h$, and with $j<1$. The divergence of the derivative of the FSC at $j = \pm 1$ is then easily gleaned
as $|1-j|^{-1/2}$, in lines with our previous discussion.  

\section{\label{Conclusions}Conclusions}

The notion of complexity has been an extremely popular theme of late, in the literature on quantum physics, both
in its first quantized and second quantized versions. Broadly, complexity measures the difficulty in preparing a quantum
state, starting from a given reference, via unitary transformations. 
In general, this might be challenging to compute, but in the context of quantum field 
theory, the notion of gauge-gravity duality has led several authors to attempt holographic computations of
quantum complexity. Two important notions of complexity exist in the literature, the 
circuit complexity or the NC \cite{Nielsen} and the FSC introduced in \cite{Chapman}. 
While NC relates to an optimal path in the space of unitary transformations, the FSC
relates to geodesic lengths on a parameter manifold. In general, these two might be different. 

In this paper, we have studied (equilibrium) ground state complexity of 
some one-dimensional exactly solvable spin models. For the NC, we find 
that its derivative diverges at a QPT, with a finite size scaling of $\partial{\mathcal C}_{N}/\partial\lambda
\sim N\log(N)$, with $N$ being the system size, and $\lambda$ being a system parameter. 
Further, in the thermodynamic limit, this derivative
diverges near criticality as $\partial{\mathcal C}_{N}/\partial\lambda\sim\log|\lambda - \lambda_c|$ with 
the critical value of $\lambda$ being $\lambda_c$. This latter fact was analytically proved in eq.(\ref{ncder}) for quasi-free 
fermionic systems, and should be universally true in all such models. 

On the other hand, our analytical results for special geodesics and numerical ones for more generic geodesics 
show that the scaling of the FSC with respect to a system parameter obeys the
relation $\partial{\mathcal C}_{FS}/\partial\lambda \sim |\lambda - \lambda_c|^{-1/2}$ for such quasi-free
fermionic systems, an analytical reasoning for which was given, leading to eq.(\ref{fscder}). 
This fact also follows purely from geometry and is model independent. 

We can compare these results with the other important indicators of QPTs, namely the Berry phase, the 
fidelity susceptibility \cite{Gu} and the Ricci scalar. To make a concrete comparison, we will focus on the transverse
XY model in the region $|h|<1$. 
As we have pointed out, the behaviour of the derivative of the Berry phase is similar to that of the 
derivative of the NC. The fidelity susceptibility has a finite size scaling $\sim N^2$ with $N$ being the 
system size, and in the thermodynamic limit, it scales with a system parameter 
as $\sim|\lambda - \lambda_c|^{-1}$, \cite{Venuti}. The Ricci scalar 
diverges as $\sim 1/\gamma$ near the anisotropy transition and is regular at the Ising transition. 

In continuation of our current study, understanding quench dynamics in the models that we have considered, via 
complexity should be an interesting next step. \\

\noindent
{\bf Acknowledgments}\\
We sincerely thank our anonymous referees for their suggestions on improving the manuscript. 
We would like to thank Arjun Bagchi, Joydeep Chakrabortty and Nilay Kundu for numerous encouraging 
discussions. N. J. would like to acknowledge the University Grant Commission (UGC), India, 
for providing financial support. T. S wishes to thank Pratim Roy for a useful discussion. 
The work of T. S. is supported in part by Science and Engineering
Research Board (India) via Project No. EMR/2016/008037.

\appendix

\section{}
\label{AppendixA}

In this appendix, we will, for completeness derive the formula for the NC of the XY spin chain, following \cite{Liu}. 
Consider the ground state of the one-dimensional spin-$1/2$ transverse XY model as a reference state, with 
parameters $h^{R}$, and $\gamma^{R}$,
\begin{eqnarray}
\ket{\Psi_{R}}&=&\prod\limits_{k=0}^{\pi}\bigg[\cos\left(\frac{\theta_{k}^{R}}{2}\right)\ket{0}_{k}
\ket{0}_{-k}-i\sin\left(\frac{\theta_{k}^{R}}{2}\right)\times\notag\\
& &\quad \quad \:e^{2i\phi}\ket{1}_{k}\ket{1}_{-k}\bigg]~.
\end{eqnarray}
The state orthogonal to this reference state is
\begin{eqnarray}
\ket{\Psi_{R}}_{\perp}&=&\prod\limits_{k=0}^{\pi}\bigg[i\sin\left(\frac{\theta_{k}^{R}}{2}\right)
e^{-2i\phi}\ket{0}_{k}\ket{0}_{-k}-\cos\left(\frac{\theta_{k}^{R}}{2}\right)\times\notag\\
& &\quad \quad \:\ket{1}_{k}\ket{1}_{-k}\bigg]~.
\end{eqnarray}
Similarly, we choose the ground state with parameters $h^{T}$ and $\gamma^{T}$ as a target state,
\begin{eqnarray}
 \ket{\Psi_{T}}&=&\prod\limits_{k=0}^{\pi}\bigg[\cos\left(\frac{\theta_{k}^{T}}{2}\right)\ket{0}_{k}
 \ket{0}_{-k}-i\sin\left(\frac{\theta_{k}^{T}}{2}\right)\times\notag\\
& &\quad \quad \:e^{2i\phi}\ket{1}_{k}\ket{1}_{-k}\bigg],
\end{eqnarray}
and the state orthogonal to the target state is then
\begin{eqnarray}
\ket{\Psi_{T}}_{\perp}&=&\prod\limits_{k=0}^{\pi}\bigg[i\sin\left(\frac{\theta_{k}^{T}}{2}\right)
e^{-2i\phi}\ket{0}_{k}\ket{0}_{-k}-\cos\left(\frac{\theta_{k}^{T}}{2}\right)\times\notag\\
& &\quad \quad \:\ket{1}_{k}\ket{1}_{-k}\bigg]~.
\end{eqnarray}
The target states $\ket{\Psi_{T}}$ and $\ket{\Psi_{T}}_{\perp}$ can be written in 
the basis of reference states $\ket{\Psi_{R}}$ and $\ket{\Psi_{R}}_{\perp}$ with the help of a unitary transformation $U_{k}$,
\begin{equation}
\begin{bmatrix}
|\psi_{T}\rangle\\|\psi_{T}\rangle\!_{\perp}
\end{bmatrix}
=U_{k}
\begin{bmatrix}
|\psi_{R}\rangle\\|\psi_{R}\rangle\!_{\perp}
\end{bmatrix},
\end{equation}
where the transformation matrix has the explicit form
\begin{equation}
U_{k}=
\begin{bmatrix}
\cos\left(\Delta\theta_{k}\right)& i\sin\left(\Delta\theta_{k}\right)e^{2i\phi}\\
i\sin\left(\Delta\theta_{k}\right)e^{-2i\phi}&\cos\left(\Delta\theta_{k}\right)\\
\end{bmatrix}, 
\end{equation}
with $\Delta\theta_{k}=\frac{\left(\theta_{k}^{T}-\theta_{k}^{R}\right)}{2}$. We follow Nielsen's 
geometric approach to find the optimal circuit required to achieve this unitary transformation. 
We first represent $U_{k}$ as a path ordered exponential
\begin{equation}
U_{k}(s)=\overleftarrow{\cal P}\exp\left[\int_{0}^{s}ds'\sum_{I}Y_{k}^{I}(s')O_{I}\right]~.
\label{pathorder}
\end{equation}
Here, $\overleftarrow{\cal P}$ denotes path ordering, and 
$Y_{k}^{I}(s)$ specifies a particular circuit being constructed by choosing a particular 
trajectory $s$, in the space of unitary circuits. The condition $U_{k}(s=0)=\mathbb{1}$ corresponds 
to the circuit that generates the initial reference state while $U_{k}(s=1)=U_{k}$ corresponds to the desired 
target state. Also, $O_{I}$ are the generators of $U(2)$, which are given by
\begin{equation}
O_{0}=
\begin{bmatrix}
i&0\\
0&i\\
\end{bmatrix}
,O_{1}=
\begin{bmatrix}
0&i\\
i&0\\
\end{bmatrix}
,O_{2}=
\begin{bmatrix}
0&1\\
-1&0\\
\end{bmatrix}
,O_{3}=
\begin{bmatrix}
i&0\\
0&-i\\
\end{bmatrix},
\end{equation}
with $I\in\{0,3\}$ in our case. To proceed further, we require the explicit forms of the function $Y_{k}^{I}(s)$.
This can be obtained by constructing a general unitary transformation $U_{k}(s)$ as,
\begin{equation}
U_{k}(s)=e^{i\beta}
\begin{bmatrix}
e^{-i\phi_{1}}\cos \omega&e^{-i\phi_{2}}\sin \omega \\
-e^{i\phi_{2}}\sin \omega& e^{i\phi_{1}}\cos \omega\\      
\end{bmatrix}~.  
\label{generalunitary}   
\end{equation}
Here $\beta$, $\omega$, $\phi_{1}$, and $\phi_{2}$ depend on the parameter $s$.  Using the relation
\begin{equation}
Tr[O_{a}O_{b}]=-2\delta_{ab},
\end{equation}
and by differentiating the expression in eq.(\ref{pathorder}) with respect to $s$, it is a fairly 
straightforward exercise to obtain 
\begin{equation}
Y_{k}^{I}(s)=-\frac{1}{2}Tr\left[(\partial_{s}U_{k}(s)) U_{k}^{-1}(s)O_{I}\right]~.
\end{equation}
Now, following standard procedure, we can define the cost (or length) functional for various paths,
\begin{equation}
{\cal D}[U_{k}]=\int_{0}^{1}ds\sum_{I}|Y_{k}^{I}(s)|^{2}~,
\label{CostFunction}
\end{equation}
and the minimal value of this functional gives the required Nielsen complexity,
\begin{equation}
{\cal C}_N[U_{k}]=min_{{Y_{k}^{I}(s)}}{\cal D}[U_{k}]~.
\label{CostFunction1}
\end{equation}
As mentioned in section \ref{complexity}, eq.(\ref{CostFunction}) refers to the $\kappa = 2$ cost
functions, to which we restrict ourselves in this paper. Generic cost functions, of the form
${\cal D}_{\kappa}[U_{k}]=\int_{0}^{1}ds\sum_{I}|Y_{k}^{I}(s)|^{\kappa}$, 
have also been studied in the literature and we refer the reader to \cite{Myers1},\cite{Myers2},\cite{Myers3}
for detailed commentary on these. 

Now, the length functional obtained for the one-dimensional spin-$1/2$ XY model is 
\begin{eqnarray}
{\cal D}[U_{k}]&=&\int_{0}^{1}ds\Bigg[\left(\frac{d\beta}{ds}\right)^{2}
+\left(\frac{d\omega}{ds}\right)^{2}+\cos^{2}\omega\left(\frac{d\phi_{1}}{ds}\right)^{2}\notag\\
&   &+\sin^{2}\omega\left(\frac{d\phi_{2}}{ds}\right)^{2}\Bigg]~.\label{integrand}
\end{eqnarray}
The above integrand in eq.(\ref{integrand}) can be minimized by setting \(\beta(s)=\phi_{1}(s)=0\), \(\phi_{2}(s)=-2\phi-\frac{\pi}{2}\), 
and \(\omega(s)=s\Delta\theta_{k}\). The NC of the desired target state is then finally obtained as,
\begin{equation}
{\cal C}_N\left(\ket{\Psi_{R}}\rightarrow\ket{\Psi_{T}}\right)=\sum\limits_{k=0}^{\pi}|\Delta\theta_{k}|^{2}~.
\label{genericParaj}
\end{equation} 
This is the generic formula for quadratic Hamiltonians, given in eq.(\ref{NC}), and 
for the transverse XY spin chain, it can be written as 
\onecolumngrid
\begin{equation}
|\Delta\theta_{k}|=\frac{1}{2}\big|\arccos\left[\frac{\left(h^{R}-\cos k\right)\left(h^{T}-\cos k\right)
+|\gamma^{R}||\gamma^{T}|(\sin k)^{2}}{\sqrt{(h^{R}-\cos k)^{2}+(\gamma^{R}\sin k)^{2}}
\sqrt{(h^{T}-\cos k)^{2}+(\gamma^{T}\sin k)^{2}}}\right]\big|~.
\label{NielsenComplexity}
\end{equation}
\twocolumngrid

\section{}
\label{AppendixB}

In this appendix, we derive the NC for the compass model of eq.(\ref{compassH}). 
The Hamiltonian obtained after a rotation by $\phi$ around $z$-direction can be expressed as 
\begin{equation}
H(\phi)= \sum\limits_{k>0}\Gamma_{k}^{\dagger}M_{k}(\phi)\Gamma_{k},
\end{equation}
where $k$ denotes the Fourier label, 
and $M_{k}(\phi)$ is a Hermitian matrix that depends on the rotation angle $\phi$, given by
\begin{equation}
M_{k}(\phi)=
\begin{bmatrix}
2\lambda&0&-A_{k}^{*}&A_{k}^{*}e^{-2i\phi}\\
0&-2\lambda&-A_{k}^{*}e^{2i\phi}&A_{k}^{*}\\
-A_{k}&-A_{k}e^{-2i\phi}&2\lambda&0\\
A_{k}e^{2i\phi}&A_{k}&0&-2\lambda
\end{bmatrix},\label{hermitianmatrix}
\end{equation}
with $\Gamma_{k}^{\dagger}=\left(a_{k}^{\dagger},a_{-k},b_{k}^{\dagger},b_{-k}\right)$,
and $A_{k}=J_{1}e^{ik}+J_{2}e^{-ik}$, and a star denoting complex conjugation. The eigenvalues of the matrix $M_{k}(\phi)$ are 
\begin{eqnarray}
\xi_{k,1}&=&-|A_{k}|-\sqrt{4\lambda^{2}+|A_{k}|^{2}},\notag\\
\xi_{k,2}&=&-|A_{k}|+\sqrt{4\lambda^{2}+|A_{k}|^{2}},\notag\\
\xi_{k,3}&=&|A_{k}|-\sqrt{4\lambda^{2}+|A_{k}|^{2}},\notag\\
\xi_{k,4}&=&|A_{k}|+\sqrt{4\lambda^{2}+|A_{k}|^{2}},
\label{engap}
\end{eqnarray}
with $|A_{k}|=\sqrt{J_{1}^{2}+J_{2}^{2}+2J_{1}J_{2}\cos(2k)}$. The above Hermitian matrix 
in eq.(\ref{hermitianmatrix}) can be diagonalized by a unitary transformation $u_{k}$ such that,
\begin{equation}
H(\phi)=\sum\limits_{k>0}\Gamma_{k}^{\dagger}u_{k}^{\dagger}u_{k}M_{k}
(\phi)u_{k}^{\dagger}u_{k}\Gamma_{k}=\sum\limits_{k>0}\Gamma_{k}^{\prime\dagger}M_{k}^{\prime}(\phi)\Gamma_{k}^{\prime},
\end{equation}
where $M_{k}^{\prime}(\phi)=u_{k}M_{k}(\phi)u_{k}^{\dagger}$ and $\Gamma_{k}^{\prime}=u_{k}\Gamma_{k}$. 
The explicit form of $u_{k}$ turns out to be 
\onecolumngrid
\begin{eqnarray}
&u_{k}=
\frac{1}{\sqrt{2}}
&\begin{bmatrix}
e^{i\phi_{k}^{A}}e^{2i\phi}\sin\frac{\theta_{k}}{2}&	e^{i\phi_{k}^{A}}\cos\frac{\theta_{k}}{2}&e^{2i\phi}
\sin\frac{\theta_{k}}{2}&-\cos\frac{\theta_{k}}{2}\\
\cos\frac{\theta_{k}}{2}&-e^{-2i\phi}\sin\frac{\theta_{k}}{2}&e^{-i\phi_{k}^{A}}\cos
\frac{\theta_{k}}{2}&e^{-i\phi_{k}^{A}}e^{-2i\phi}\sin\frac{\theta_{k}}{2}\\
-e^{i\phi_{k}^{A}}e^{2i\phi}\sin\frac{\theta_{k}}{2}&e^{i\phi_{k}^{A}}\cos\frac{\theta_{k}}{2}&e^{2i\phi}
\sin\frac{\theta_{k}}{2}&\cos\frac{\theta_{k}}{2}\\
-\cos\frac{\theta_{k}}{2}&-e^{-2i\phi}\sin\frac{\theta_{k}}{2}&e^{-i\phi_{k}^{A}}
\cos\frac{\theta_{k}}{2}&-e^{-i\phi_{k}^{A}}e^{-2i\phi}\sin\frac{\theta_{k}}{2}
\end{bmatrix},
\end{eqnarray}
\twocolumngrid
\noindent
with $\phi_{k}^{A}=\text{arg} (A_{k})$ and $\cos\theta_{k}=2\lambda/\sqrt{4\lambda^{2}+|A_{k}|^{2}}$.
Now, the ground state energy is given by
\begin{equation}
\xi_{k,1}=-|A_{k}|-\sqrt{4\lambda^{2}+|A_{k}|^{2}}~.
\end{equation}
The eigenvector corresponding to the ground state energy $\xi_{k,1}$ is given by,
\begin{equation}
\ket{g_{k}}=\frac{1}{\sqrt{2}}
\begin{bmatrix}
-e^{-2i\phi}e^{-i\phi_{k}^{A}}\sin\left(\frac{\theta_{k}}{2}\right)\\
-e^{-i\phi_{k}^{A}}\cos\left(\frac{\theta_{k}}{2}\right)\\
-e^{-2i\phi}\sin\left(\frac{\theta_{k}}{2}\right)\\
\cos\left(\frac{\theta_{k}}{2}\right)
\end{bmatrix}~.
\end{equation}
The NC of the desired target state is again obtained as,
\begin{equation}
{\cal C}_N\left(\ket{\Psi_{R}}\rightarrow\ket{\Psi_{T}}\right)=\sum\limits_{k=0}^{\pi}|\Delta\theta_{k}|^{2},
\end{equation}
where we have with $J_1=1$,
\begin{eqnarray}
|\Delta\theta_{k}|&=&\frac{1}{2}\big|\arccos\left[\frac{4\lambda^{R}\lambda^{T}+|A_{k}^{R}|
|A_{k}^{T}|}{\sqrt{4\left(\lambda^{R}\right)^{2}+|A_{k}^{R}|^{2}}\sqrt{4\left(\lambda^{T}\right)^{2}+|A_{k}^{T}|^{2}}}\right]\big|~,\nonumber\\
|A_{k}^{R}|&=&\sqrt{1+\left(J_{2}^{R}\right)^{2}+2J_{2}^{R}\cos(2k)},\notag\\
|A_{k}^{T}|&=&\sqrt{1+\left(J_{2}^{T}\right)^{2}+2J_{2}^{T}\cos(2k)}~.
\label{NCcompassH}
\end{eqnarray}
\\
\bibliographystyle{apsrev4-1}
\bibliography{bibfile}

\begin{thebibliography}{42}%
\makeatletter
\providecommand \@ifxundefined [1]{%
 \@ifx{#1\undefined}
}%
\providecommand \@ifnum [1]{%
 \ifnum #1\expandafter \@firstoftwo
 \else \expandafter \@secondoftwo
 \fi
}%
\providecommand \@ifx [1]{%
 \ifx #1\expandafter \@firstoftwo
 \else \expandafter \@secondoftwo
 \fi
}%
\providecommand \natexlab [1]{#1}%
\providecommand \enquote  [1]{``#1''}%
\providecommand \bibnamefont  [1]{#1}%
\providecommand \bibfnamefont [1]{#1}%
\providecommand \citenamefont [1]{#1}%
\providecommand \href@noop [0]{\@secondoftwo}%
\providecommand \href [0]{\begingroup \@sanitize@url \@href}%
\providecommand \@href[1]{\@@startlink{#1}\@@href}%
\providecommand \@@href[1]{\endgroup#1\@@endlink}%
\providecommand \@sanitize@url [0]{\catcode `\\12\catcode `\$12\catcode
  `\&12\catcode `\#12\catcode `\^12\catcode `\_12\catcode `\%12\relax}%
\providecommand \@@startlink[1]{}%
\providecommand \@@endlink[0]{}%
\providecommand \url  [0]{\begingroup\@sanitize@url \@url }%
\providecommand \@url [1]{\endgroup\@href {#1}{\urlprefix }}%
\providecommand \urlprefix  [0]{URL }%
\providecommand \Eprint [0]{\href }%
\providecommand \doibase [0]{http://dx.doi.org/}%
\providecommand \selectlanguage [0]{\@gobble}%
\providecommand \bibinfo  [0]{\@secondoftwo}%
\providecommand \bibfield  [0]{\@secondoftwo}%
\providecommand \translation [1]{[#1]}%
\providecommand \BibitemOpen [0]{}%
\providecommand \bibitemStop [0]{}%
\providecommand \bibitemNoStop [0]{.\EOS\space}%
\providecommand \EOS [0]{\spacefactor3000\relax}%
\providecommand \BibitemShut  [1]{\csname bibitem#1\endcsname}%
\let\auto@bib@innerbib\@empty
\bibitem [{\citenamefont {Sachdev}(2011)}]{Subir}%
  \BibitemOpen
  \bibfield  {author} {\bibinfo {author} {\bibfnamefont {S.}~\bibnamefont
  {Sachdev}},\ }\href {\doibase 10.1017/CBO9780511973765} {\emph {\bibinfo
  {title} {Quantum Phase Transitions}}},\ \bibinfo {edition} {2nd}\ ed.\
  (\bibinfo  {publisher} {Cambridge University Press, Cambridge},\ \bibinfo
  {year} {2011})\BibitemShut {NoStop}%
\bibitem [{\citenamefont {Goldenfeld}(2018)}]{Nigel}%
  \BibitemOpen
  \bibfield  {author} {\bibinfo {author} {\bibfnamefont {N.}~\bibnamefont
  {Goldenfeld}},\ }\href@noop {} {\emph {\bibinfo {title} {Lectures on phase
  transitions and the renormalization group}}}\ (\bibinfo  {publisher} {CRC
  Press},\ \bibinfo {year} {2018})\BibitemShut {NoStop}%
\bibitem [{\citenamefont {Cardy}(1996)}]{Cardy}%
  \BibitemOpen
  \bibfield  {author} {\bibinfo {author} {\bibfnamefont {J.}~\bibnamefont
  {Cardy}},\ }\href@noop {} {\emph {\bibinfo {title} {Scaling and
  renormalization in statistical physics}}}\ (\bibinfo  {publisher} {Cambridge
  University Press},\ \bibinfo {year} {1996})\BibitemShut {NoStop}%
\bibitem [{\citenamefont {Stanley}(1971)}]{Stanley}%
  \BibitemOpen
  \bibfield  {author} {\bibinfo {author} {\bibfnamefont {H.~E.}\ \bibnamefont
  {Stanley}},\ }\href@noop {} {\emph {\bibinfo {title} {Phase transitions and
  critical phenomena}}}\ (\bibinfo  {publisher} {Clarendon Press, Oxford},\
  \bibinfo {year} {1971})\BibitemShut {NoStop}%
\bibitem [{\citenamefont {Brody}\ and\ \citenamefont {Hook}(2008)}]{BrodyHook}%
  \BibitemOpen
  \bibfield  {author} {\bibinfo {author} {\bibfnamefont {D.~C.}\ \bibnamefont
  {Brody}}\ and\ \bibinfo {author} {\bibfnamefont {D.~W.}\ \bibnamefont
  {Hook}},\ }\href {\doibase 10.1088/1751-8113/42/2/023001} {\bibfield
  {journal} {\bibinfo  {journal} {Journal of Physics A: Mathematical and
  Theoretical}\ }\textbf {\bibinfo {volume} {42}},\ \bibinfo {pages} {023001}
  (\bibinfo {year} {2008})}\BibitemShut {NoStop}%
\bibitem [{\citenamefont {Ruppeiner}(1995)}]{Ruppeiner}%
  \BibitemOpen
  \bibfield  {author} {\bibinfo {author} {\bibfnamefont {G.}~\bibnamefont
  {Ruppeiner}},\ }\href {\doibase 10.1103/RevModPhys.67.605} {\bibfield
  {journal} {\bibinfo  {journal} {Rev. Mod. Phys.}\ }\textbf {\bibinfo {volume}
  {67}},\ \bibinfo {pages} {605} (\bibinfo {year} {1995})},\ \bibinfo {note}
  {[Erratum: Rev. Mod. Phys.68,313(1996)]}\BibitemShut {NoStop}%
\bibitem [{\citenamefont {Provost}\ and\ \citenamefont
  {Vallee}(1980)}]{Provost}%
  \BibitemOpen
  \bibfield  {author} {\bibinfo {author} {\bibfnamefont {J.}~\bibnamefont
  {Provost}}\ and\ \bibinfo {author} {\bibfnamefont {G.}~\bibnamefont
  {Vallee}},\ }\href@noop {} {\ \textbf {\bibinfo {volume} {76}},\ \bibinfo
  {pages} {289} (\bibinfo {year} {1980})}\BibitemShut {NoStop}%
\bibitem [{\citenamefont {Kolodrubetz}\ \emph {et~al.}(2013)\citenamefont
  {Kolodrubetz}, \citenamefont {Gritsev},\ and\ \citenamefont
  {Polkovnikov}}]{Polkovnikov}%
  \BibitemOpen
  \bibfield  {author} {\bibinfo {author} {\bibfnamefont {M.}~\bibnamefont
  {Kolodrubetz}}, \bibinfo {author} {\bibfnamefont {V.}~\bibnamefont
  {Gritsev}}, \ and\ \bibinfo {author} {\bibfnamefont {A.}~\bibnamefont
  {Polkovnikov}},\ }\href@noop {} {\bibfield  {journal} {\bibinfo  {journal}
  {Phys. Rev. B}\ }\textbf {\bibinfo {volume} {88}},\ \bibinfo {pages} {064304}
  (\bibinfo {year} {2013})}\BibitemShut {NoStop}%
\bibitem [{\citenamefont {Zanardi}\ \emph {et~al.}(2007)\citenamefont
  {Zanardi}, \citenamefont {Giorda},\ and\ \citenamefont {Cozzini}}]{Zanardi}%
  \BibitemOpen
  \bibfield  {author} {\bibinfo {author} {\bibfnamefont {P.}~\bibnamefont
  {Zanardi}}, \bibinfo {author} {\bibfnamefont {P.}~\bibnamefont {Giorda}}, \
  and\ \bibinfo {author} {\bibfnamefont {M.}~\bibnamefont {Cozzini}},\
  }\href@noop {} {\bibfield  {journal} {\bibinfo  {journal} {Phys. Rev. Lett.}\
  }\textbf {\bibinfo {volume} {99}},\ \bibinfo {pages} {100603} (\bibinfo
  {year} {2007})}\BibitemShut {NoStop}%
\bibitem [{\citenamefont {Maity}\ \emph {et~al.}(2015)\citenamefont {Maity},
  \citenamefont {Mahapatra},\ and\ \citenamefont {Sarkar}}]{Sarkar}%
  \BibitemOpen
  \bibfield  {author} {\bibinfo {author} {\bibfnamefont {R.}~\bibnamefont
  {Maity}}, \bibinfo {author} {\bibfnamefont {S.}~\bibnamefont {Mahapatra}}, \
  and\ \bibinfo {author} {\bibfnamefont {T.}~\bibnamefont {Sarkar}},\
  }\href@noop {} {\bibfield  {journal} {\bibinfo  {journal} {Phys. Rev. E}\
  }\textbf {\bibinfo {volume} {92}},\ \bibinfo {pages} {052101} (\bibinfo
  {year} {2015})}\BibitemShut {NoStop}%
\bibitem [{\citenamefont {Dey}\ \emph {et~al.}(2012)\citenamefont {Dey},
  \citenamefont {Mahapatra}, \citenamefont {Roy},\ and\ \citenamefont
  {Sarkar}}]{TapoDickeModel}%
  \BibitemOpen
  \bibfield  {author} {\bibinfo {author} {\bibfnamefont {A.}~\bibnamefont
  {Dey}}, \bibinfo {author} {\bibfnamefont {S.}~\bibnamefont {Mahapatra}},
  \bibinfo {author} {\bibfnamefont {P.}~\bibnamefont {Roy}}, \ and\ \bibinfo
  {author} {\bibfnamefont {T.}~\bibnamefont {Sarkar}},\ }\href {\doibase
  10.1103/PhysRevE.86.031137} {\bibfield  {journal} {\bibinfo  {journal} {Phys.
  Rev.}\ }\textbf {\bibinfo {volume} {E86}},\ \bibinfo {pages} {031137}
  (\bibinfo {year} {2012})},\ \Eprint {http://arxiv.org/abs/1208.4710}
  {arXiv:1208.4710 [cond-mat.stat-mech]} \BibitemShut {NoStop}%
\bibitem [{\citenamefont {Nielsen}(2005)}]{Nielsen}%
  \BibitemOpen
  \bibfield  {author} {\bibinfo {author} {\bibfnamefont {M.~A.}\ \bibnamefont
  {Nielsen}},\ }\href@noop {} {\bibfield  {journal} {\bibinfo  {journal} {arXiv
  preprint quant-ph/0502070}\ } (\bibinfo {year} {2005})}\BibitemShut {NoStop}%
\bibitem [{\citenamefont {Liu}\ \emph {et~al.}(2020)\citenamefont {Liu},
  \citenamefont {Whitsitt}, \citenamefont {Curtis}, \citenamefont {Lundgren},
  \citenamefont {Titum}, \citenamefont {Yang}, \citenamefont {Garrison},\ and\
  \citenamefont {Gorshkov}}]{Liu}%
  \BibitemOpen
  \bibfield  {author} {\bibinfo {author} {\bibfnamefont {F.}~\bibnamefont
  {Liu}}, \bibinfo {author} {\bibfnamefont {S.}~\bibnamefont {Whitsitt}},
  \bibinfo {author} {\bibfnamefont {J.~B.}\ \bibnamefont {Curtis}}, \bibinfo
  {author} {\bibfnamefont {R.}~\bibnamefont {Lundgren}}, \bibinfo {author}
  {\bibfnamefont {P.}~\bibnamefont {Titum}}, \bibinfo {author} {\bibfnamefont
  {Z.-C.}\ \bibnamefont {Yang}}, \bibinfo {author} {\bibfnamefont {J.~R.}\
  \bibnamefont {Garrison}}, \ and\ \bibinfo {author} {\bibfnamefont {A.~V.}\
  \bibnamefont {Gorshkov}},\ }\href {\doibase 10.1103/PhysRevResearch.2.013323}
  {\bibfield  {journal} {\bibinfo  {journal} {Phys. Rev. Res.}\ }\textbf
  {\bibinfo {volume} {2}},\ \bibinfo {pages} {013323} (\bibinfo {year}
  {2020})},\ \bibinfo {note} {[Phys. Rev. Research.2,013323(2020)]},\ \Eprint
  {http://arxiv.org/abs/1902.10720} {arXiv:1902.10720 [quant-ph]} \BibitemShut
  {NoStop}%
\bibitem [{\citenamefont {Khan}\ \emph {et~al.}(2018)\citenamefont {Khan},
  \citenamefont {Krishnan},\ and\ \citenamefont {Sharma}}]{Khan}%
  \BibitemOpen
  \bibfield  {author} {\bibinfo {author} {\bibfnamefont {R.}~\bibnamefont
  {Khan}}, \bibinfo {author} {\bibfnamefont {C.}~\bibnamefont {Krishnan}}, \
  and\ \bibinfo {author} {\bibfnamefont {S.}~\bibnamefont {Sharma}},\
  }\href@noop {} {\bibfield  {journal} {\bibinfo  {journal} {Phys. Rev. D}\
  }\textbf {\bibinfo {volume} {98}},\ \bibinfo {pages} {126001} (\bibinfo
  {year} {2018})}\BibitemShut {NoStop}%
\bibitem [{\citenamefont {Xiong}\ \emph {et~al.}(2020)\citenamefont {Xiong},
  \citenamefont {Yao},\ and\ \citenamefont {Yan}}]{Xiong}%
  \BibitemOpen
  \bibfield  {author} {\bibinfo {author} {\bibfnamefont {Z.}~\bibnamefont
  {Xiong}}, \bibinfo {author} {\bibfnamefont {D.-X.}\ \bibnamefont {Yao}}, \
  and\ \bibinfo {author} {\bibfnamefont {Z.}~\bibnamefont {Yan}},\ }\href
  {\doibase 10.1103/PhysRevB.101.174305} {\bibfield  {journal} {\bibinfo
  {journal} {Phys. Rev. B}\ }\textbf {\bibinfo {volume} {101}},\ \bibinfo
  {pages} {174305} (\bibinfo {year} {2020})}\BibitemShut {NoStop}%
\bibitem [{\citenamefont {Jefferson}\ and\ \citenamefont
  {Myers}(2017)}]{Myers1}%
  \BibitemOpen
  \bibfield  {author} {\bibinfo {author} {\bibfnamefont {R.}~\bibnamefont
  {Jefferson}}\ and\ \bibinfo {author} {\bibfnamefont {R.~C.}\ \bibnamefont
  {Myers}},\ }\href {\doibase 10.1007/JHEP10(2017)107} {\bibfield  {journal}
  {\bibinfo  {journal} {JHEP}\ }\textbf {\bibinfo {volume} {10}},\ \bibinfo
  {pages} {107} (\bibinfo {year} {2017})},\ \Eprint
  {http://arxiv.org/abs/1707.08570} {arXiv:1707.08570 [hep-th]} \BibitemShut
  {NoStop}%
\bibitem [{\citenamefont {Guo}\ \emph {et~al.}(2018)\citenamefont {Guo},
  \citenamefont {Hernandez}, \citenamefont {Myers},\ and\ \citenamefont
  {Ruan}}]{Myers2}%
  \BibitemOpen
  \bibfield  {author} {\bibinfo {author} {\bibfnamefont {M.}~\bibnamefont
  {Guo}}, \bibinfo {author} {\bibfnamefont {J.}~\bibnamefont {Hernandez}},
  \bibinfo {author} {\bibfnamefont {R.~C.}\ \bibnamefont {Myers}}, \ and\
  \bibinfo {author} {\bibfnamefont {S.-M.}\ \bibnamefont {Ruan}},\ }\href
  {\doibase 10.1007/JHEP10(2018)011} {\bibfield  {journal} {\bibinfo  {journal}
  {JHEP}\ }\textbf {\bibinfo {volume} {10}},\ \bibinfo {pages} {011} (\bibinfo
  {year} {2018})},\ \Eprint {http://arxiv.org/abs/1807.07677} {arXiv:1807.07677
  [hep-th]} \BibitemShut {NoStop}%
\bibitem [{\citenamefont {Camargo}\ \emph {et~al.}(2019)\citenamefont
  {Camargo}, \citenamefont {Caputa}, \citenamefont {Das}, \citenamefont
  {Heller},\ and\ \citenamefont {Jefferson}}]{Diptarka}%
  \BibitemOpen
  \bibfield  {author} {\bibinfo {author} {\bibfnamefont {H.~A.}\ \bibnamefont
  {Camargo}}, \bibinfo {author} {\bibfnamefont {P.}~\bibnamefont {Caputa}},
  \bibinfo {author} {\bibfnamefont {D.}~\bibnamefont {Das}}, \bibinfo {author}
  {\bibfnamefont {M.~P.}\ \bibnamefont {Heller}}, \ and\ \bibinfo {author}
  {\bibfnamefont {R.}~\bibnamefont {Jefferson}},\ }\href {\doibase
  10.1103/PhysRevLett.122.081601} {\bibfield  {journal} {\bibinfo  {journal}
  {Phys. Rev. Lett.}\ }\textbf {\bibinfo {volume} {122}},\ \bibinfo {pages}
  {081601} (\bibinfo {year} {2019})},\ \Eprint
  {http://arxiv.org/abs/1807.07075} {arXiv:1807.07075 [hep-th]} \BibitemShut
  {NoStop}%
\bibitem [{\citenamefont {Ali}\ \emph {et~al.}(2018)\citenamefont {Ali},
  \citenamefont {Bhattacharyya}, \citenamefont {Shajidul~Haque}, \citenamefont
  {Kim},\ and\ \citenamefont {Moynihan}}]{Arpan}%
  \BibitemOpen
  \bibfield  {author} {\bibinfo {author} {\bibfnamefont {T.}~\bibnamefont
  {Ali}}, \bibinfo {author} {\bibfnamefont {A.}~\bibnamefont {Bhattacharyya}},
  \bibinfo {author} {\bibfnamefont {S.}~\bibnamefont {Shajidul~Haque}},
  \bibinfo {author} {\bibfnamefont {E.~H.}\ \bibnamefont {Kim}}, \ and\
  \bibinfo {author} {\bibfnamefont {N.}~\bibnamefont {Moynihan}},\ }\href@noop
  {} {\  (\bibinfo {year} {2018})},\ \Eprint {http://arxiv.org/abs/1811.05985}
  {arXiv:1811.05985 [hep-th]} \BibitemShut {NoStop}%
\bibitem [{\citenamefont {Bhattacharyya}\ \emph {et~al.}(2018)\citenamefont
  {Bhattacharyya}, \citenamefont {Shekar},\ and\ \citenamefont
  {Sinha}}]{Arpan2}%
  \BibitemOpen
  \bibfield  {author} {\bibinfo {author} {\bibfnamefont {A.}~\bibnamefont
  {Bhattacharyya}}, \bibinfo {author} {\bibfnamefont {A.}~\bibnamefont
  {Shekar}}, \ and\ \bibinfo {author} {\bibfnamefont {A.}~\bibnamefont
  {Sinha}},\ }\href {\doibase 10.1007/JHEP10(2018)140} {\bibfield  {journal}
  {\bibinfo  {journal} {JHEP}\ }\textbf {\bibinfo {volume} {10}},\ \bibinfo
  {pages} {140} (\bibinfo {year} {2018})},\ \Eprint
  {http://arxiv.org/abs/1808.03105} {arXiv:1808.03105 [hep-th]} \BibitemShut
  {NoStop}%
\bibitem [{\citenamefont {Guo}\ \emph {et~al.}(2020)\citenamefont {Guo},
  \citenamefont {Fan}, \citenamefont {Jiang}, \citenamefont {Liu},\ and\
  \citenamefont {Chen}}]{Guo}%
  \BibitemOpen
  \bibfield  {author} {\bibinfo {author} {\bibfnamefont {M.}~\bibnamefont
  {Guo}}, \bibinfo {author} {\bibfnamefont {Z.-Y.}\ \bibnamefont {Fan}},
  \bibinfo {author} {\bibfnamefont {J.}~\bibnamefont {Jiang}}, \bibinfo
  {author} {\bibfnamefont {X.}~\bibnamefont {Liu}}, \ and\ \bibinfo {author}
  {\bibfnamefont {B.}~\bibnamefont {Chen}},\ }\href@noop {} {\  (\bibinfo
  {year} {2020})},\ \Eprint {http://arxiv.org/abs/2004.00344} {arXiv:2004.00344
  [hep-th]} \BibitemShut {NoStop}%
\bibitem [{\citenamefont {Chapman}\ \emph {et~al.}(2018)\citenamefont
  {Chapman}, \citenamefont {Heller}, \citenamefont {Marrochio},\ and\
  \citenamefont {Pastawski}}]{Chapman}%
  \BibitemOpen
  \bibfield  {author} {\bibinfo {author} {\bibfnamefont {S.}~\bibnamefont
  {Chapman}}, \bibinfo {author} {\bibfnamefont {M.~P.}\ \bibnamefont {Heller}},
  \bibinfo {author} {\bibfnamefont {H.}~\bibnamefont {Marrochio}}, \ and\
  \bibinfo {author} {\bibfnamefont {F.}~\bibnamefont {Pastawski}},\ }\href
  {\doibase 10.1103/PhysRevLett.120.121602} {\bibfield  {journal} {\bibinfo
  {journal} {Phys. Rev. Lett.}\ }\textbf {\bibinfo {volume} {120}},\ \bibinfo
  {pages} {121602} (\bibinfo {year} {2018})},\ \Eprint
  {http://arxiv.org/abs/1707.08582} {arXiv:1707.08582 [hep-th]} \BibitemShut
  {NoStop}%
\bibitem [{\citenamefont {Hackl}\ and\ \citenamefont {Myers}(2018)}]{Myers3}%
  \BibitemOpen
  \bibfield  {author} {\bibinfo {author} {\bibfnamefont {L.}~\bibnamefont
  {Hackl}}\ and\ \bibinfo {author} {\bibfnamefont {R.~C.}\ \bibnamefont
  {Myers}},\ }\href {\doibase 10.1007/JHEP07(2018)139} {\bibfield  {journal}
  {\bibinfo  {journal} {JHEP}\ }\textbf {\bibinfo {volume} {07}},\ \bibinfo
  {pages} {139} (\bibinfo {year} {2018})},\ \Eprint
  {http://arxiv.org/abs/1803.10638} {arXiv:1803.10638 [hep-th]} \BibitemShut
  {NoStop}%
\bibitem [{\citenamefont {Kumar}\ and\ \citenamefont
  {Sarkar}(2014)}]{TapoKumar}%
  \BibitemOpen
  \bibfield  {author} {\bibinfo {author} {\bibfnamefont {P.}~\bibnamefont
  {Kumar}}\ and\ \bibinfo {author} {\bibfnamefont {T.}~\bibnamefont {Sarkar}},\
  }\href {\doibase 10.1103/PhysRevE.90.042145} {\bibfield  {journal} {\bibinfo
  {journal} {Phys. Rev.}\ }\textbf {\bibinfo {volume} {E90}},\ \bibinfo {pages}
  {042145} (\bibinfo {year} {2014})},\ \Eprint {http://arxiv.org/abs/1405.3212}
  {arXiv:1405.3212 [cond-mat.stat-mech]} \BibitemShut {NoStop}%
\bibitem [{\citenamefont {Lieb}\ \emph {et~al.}(1961)\citenamefont {Lieb},
  \citenamefont {Schultz},\ and\ \citenamefont {Mattis}}]{LSM}%
  \BibitemOpen
  \bibfield  {author} {\bibinfo {author} {\bibfnamefont {E.}~\bibnamefont
  {Lieb}}, \bibinfo {author} {\bibfnamefont {T.}~\bibnamefont {Schultz}}, \
  and\ \bibinfo {author} {\bibfnamefont {D.}~\bibnamefont {Mattis}},\
  }\href@noop {} {\bibfield  {journal} {\bibinfo  {journal} {Annals of
  Physics}\ }\textbf {\bibinfo {volume} {16}},\ \bibinfo {pages} {407}
  (\bibinfo {year} {1961})}\BibitemShut {NoStop}%
\bibitem [{\citenamefont {Katsura}(1962)}]{Katsura1}%
  \BibitemOpen
  \bibfield  {author} {\bibinfo {author} {\bibfnamefont {S.}~\bibnamefont
  {Katsura}},\ }\href {\doibase 10.1103/PhysRev.127.1508} {\bibfield  {journal}
  {\bibinfo  {journal} {Phys. Rev.}\ }\textbf {\bibinfo {volume} {127}},\
  \bibinfo {pages} {1508} (\bibinfo {year} {1962})}\BibitemShut {NoStop}%
\bibitem [{\citenamefont {Barouch}\ and\ \citenamefont
  {McCoy}(1971)}]{BarouchMcCoy}%
  \BibitemOpen
  \bibfield  {author} {\bibinfo {author} {\bibfnamefont {E.}~\bibnamefont
  {Barouch}}\ and\ \bibinfo {author} {\bibfnamefont {B.~M.}\ \bibnamefont
  {McCoy}},\ }\href {\doibase 10.1103/PhysRevA.3.786} {\bibfield  {journal}
  {\bibinfo  {journal} {Phys. Rev. A}\ }\textbf {\bibinfo {volume} {3}},\
  \bibinfo {pages} {786} (\bibinfo {year} {1971})}\BibitemShut {NoStop}%
\bibitem [{\citenamefont {Bunder}\ and\ \citenamefont
  {McKenzie}(1999)}]{Bunder}%
  \BibitemOpen
  \bibfield  {author} {\bibinfo {author} {\bibfnamefont {J.~E.}\ \bibnamefont
  {Bunder}}\ and\ \bibinfo {author} {\bibfnamefont {R.~H.}\ \bibnamefont
  {McKenzie}},\ }\href {\doibase 10.1103/PhysRevB.60.344} {\bibfield  {journal}
  {\bibinfo  {journal} {Phys. Rev. B}\ }\textbf {\bibinfo {volume} {60}},\
  \bibinfo {pages} {344} (\bibinfo {year} {1999})}\BibitemShut {NoStop}%
\bibitem [{\citenamefont {Mukherjee}\ \emph {et~al.}(2007)\citenamefont
  {Mukherjee}, \citenamefont {Divakaran}, \citenamefont {Dutta},\ and\
  \citenamefont {Sen}}]{Amit}%
  \BibitemOpen
  \bibfield  {author} {\bibinfo {author} {\bibfnamefont {V.}~\bibnamefont
  {Mukherjee}}, \bibinfo {author} {\bibfnamefont {U.}~\bibnamefont
  {Divakaran}}, \bibinfo {author} {\bibfnamefont {A.}~\bibnamefont {Dutta}}, \
  and\ \bibinfo {author} {\bibfnamefont {D.}~\bibnamefont {Sen}},\ }\href
  {\doibase 10.1103/PhysRevB.76.174303} {\bibfield  {journal} {\bibinfo
  {journal} {Phys. Rev. B}\ }\textbf {\bibinfo {volume} {76}},\ \bibinfo
  {pages} {174303} (\bibinfo {year} {2007})}\BibitemShut {NoStop}%
\bibitem [{\citenamefont {Dutta}\ \emph {et~al.}(2015)\citenamefont {Dutta},
  \citenamefont {Aepplie}, \citenamefont {Chakrabarti}, \citenamefont {Uma},
  \citenamefont {Rosenbaum},\ and\ \citenamefont {Sen}}]{AmitBook}%
  \BibitemOpen
  \bibfield  {author} {\bibinfo {author} {\bibfnamefont {A.}~\bibnamefont
  {Dutta}}, \bibinfo {author} {\bibfnamefont {G.}~\bibnamefont {Aepplie}},
  \bibinfo {author} {\bibfnamefont {B.~K.}\ \bibnamefont {Chakrabarti}},
  \bibinfo {author} {\bibfnamefont {D.}~\bibnamefont {Uma}}, \bibinfo {author}
  {\bibfnamefont {T.~F.}\ \bibnamefont {Rosenbaum}}, \ and\ \bibinfo {author}
  {\bibfnamefont {D.}~\bibnamefont {Sen}},\ }\href@noop {} {\emph {\bibinfo
  {title} {Quantum Phase Transitions in Transverse Field Models}}}\ (\bibinfo
  {publisher} {Cambridge University Press},\ \bibinfo {year}
  {2015})\BibitemShut {NoStop}%
\bibitem [{\citenamefont {Zhu}(2006)}]{Zhu}%
  \BibitemOpen
  \bibfield  {author} {\bibinfo {author} {\bibfnamefont {S.-L.}\ \bibnamefont
  {Zhu}},\ }\href@noop {} {\bibfield  {journal} {\bibinfo  {journal} {Phys.
  Rev. Lett.}\ }\textbf {\bibinfo {volume} {96}},\ \bibinfo {pages} {077206}
  (\bibinfo {year} {2006})}\BibitemShut {NoStop}%
\bibitem [{\citenamefont {Venuti}\ and\ \citenamefont
  {Zanardi}(2007)}]{Venuti}%
  \BibitemOpen
  \bibfield  {author} {\bibinfo {author} {\bibfnamefont {L.~C.}\ \bibnamefont
  {Venuti}}\ and\ \bibinfo {author} {\bibfnamefont {P.}~\bibnamefont
  {Zanardi}},\ }\href@noop {} {\bibfield  {journal} {\bibinfo  {journal} {Phys.
  Rev. Lett.}\ }\textbf {\bibinfo {volume} {99}},\ \bibinfo {pages} {095701}
  (\bibinfo {year} {2007})}\BibitemShut {NoStop}%
\bibitem [{\citenamefont {Kumar}\ \emph {et~al.}(2012)\citenamefont {Kumar},
  \citenamefont {Mahapatra}, \citenamefont {Phukon},\ and\ \citenamefont
  {Sarkar}}]{TapoGeodesics}%
  \BibitemOpen
  \bibfield  {author} {\bibinfo {author} {\bibfnamefont {P.}~\bibnamefont
  {Kumar}}, \bibinfo {author} {\bibfnamefont {S.}~\bibnamefont {Mahapatra}},
  \bibinfo {author} {\bibfnamefont {P.}~\bibnamefont {Phukon}}, \ and\ \bibinfo
  {author} {\bibfnamefont {T.}~\bibnamefont {Sarkar}},\ }\href {\doibase
  10.1103/PhysRevE.86.051117} {\bibfield  {journal} {\bibinfo  {journal} {Phys.
  Rev.}\ }\textbf {\bibinfo {volume} {E86}},\ \bibinfo {pages} {051117}
  (\bibinfo {year} {2012})},\ \Eprint {http://arxiv.org/abs/1210.7135}
  {arXiv:1210.7135 [cond-mat.stat-mech]} \BibitemShut {NoStop}%
\bibitem [{\citenamefont {Perk}\ \emph {et~al.}(1975)\citenamefont {Perk},
  \citenamefont {Capel}, \citenamefont {Zuilhof},\ and\ \citenamefont
  {Siskens}}]{Perk}%
  \BibitemOpen
  \bibfield  {author} {\bibinfo {author} {\bibfnamefont {J.}~\bibnamefont
  {Perk}}, \bibinfo {author} {\bibfnamefont {H.}~\bibnamefont {Capel}},
  \bibinfo {author} {\bibfnamefont {M.}~\bibnamefont {Zuilhof}}, \ and\
  \bibinfo {author} {\bibfnamefont {T.~J.}\ \bibnamefont {Siskens}},\
  }\href@noop {} {\bibfield  {journal} {\bibinfo  {journal} {Physica A:
  Statistical Mechanics and its Applications}\ }\textbf {\bibinfo {volume}
  {81}},\ \bibinfo {pages} {319} (\bibinfo {year} {1975})}\BibitemShut
  {NoStop}%
\bibitem [{\citenamefont {Brzezicki}\ \emph {et~al.}(2007)\citenamefont
  {Brzezicki}, \citenamefont {Dziarmaga},\ and\ \citenamefont
  {Ole\ifmmode~\acute{s}\else \'{s}\fi{}}}]{Compass0}%
  \BibitemOpen
  \bibfield  {author} {\bibinfo {author} {\bibfnamefont {W.}~\bibnamefont
  {Brzezicki}}, \bibinfo {author} {\bibfnamefont {J.}~\bibnamefont
  {Dziarmaga}}, \ and\ \bibinfo {author} {\bibfnamefont {A.~M.}\ \bibnamefont
  {Ole\ifmmode~\acute{s}\else \'{s}\fi{}}},\ }\href {\doibase
  10.1103/PhysRevB.75.134415} {\bibfield  {journal} {\bibinfo  {journal} {Phys.
  Rev. B}\ }\textbf {\bibinfo {volume} {75}},\ \bibinfo {pages} {134415}
  (\bibinfo {year} {2007})}\BibitemShut {NoStop}%
\bibitem [{\citenamefont {You}\ and\ \citenamefont {Tian}(2008)}]{Compass1}%
  \BibitemOpen
  \bibfield  {author} {\bibinfo {author} {\bibfnamefont {W.-L.}\ \bibnamefont
  {You}}\ and\ \bibinfo {author} {\bibfnamefont {G.-S.}\ \bibnamefont {Tian}},\
  }\href {\doibase 10.1103/PhysRevB.78.184406} {\bibfield  {journal} {\bibinfo
  {journal} {Phys. Rev. B}\ }\textbf {\bibinfo {volume} {78}},\ \bibinfo
  {pages} {184406} (\bibinfo {year} {2008})}\BibitemShut {NoStop}%
\bibitem [{\citenamefont {Eriksson}\ and\ \citenamefont
  {Johannesson}(2009)}]{Compass2}%
  \BibitemOpen
  \bibfield  {author} {\bibinfo {author} {\bibfnamefont {E.}~\bibnamefont
  {Eriksson}}\ and\ \bibinfo {author} {\bibfnamefont {H.}~\bibnamefont
  {Johannesson}},\ }\href {\doibase 10.1103/PhysRevB.79.224424} {\bibfield
  {journal} {\bibinfo  {journal} {Phys. Rev. B}\ }\textbf {\bibinfo {volume}
  {79}},\ \bibinfo {pages} {224424} (\bibinfo {year} {2009})}\BibitemShut
  {NoStop}%
\bibitem [{\citenamefont {Jafari}\ and\ \citenamefont
  {Johannesson}(2017)}]{Jaffari1}%
  \BibitemOpen
  \bibfield  {author} {\bibinfo {author} {\bibfnamefont {R.}~\bibnamefont
  {Jafari}}\ and\ \bibinfo {author} {\bibfnamefont {H.}~\bibnamefont
  {Johannesson}},\ }\href {\doibase 10.1103/PhysRevB.96.224302} {\bibfield
  {journal} {\bibinfo  {journal} {Phys. Rev. B}\ }\textbf {\bibinfo {volume}
  {96}},\ \bibinfo {pages} {224302} (\bibinfo {year} {2017})}\BibitemShut
  {NoStop}%
\bibitem [{\citenamefont {Jafari}(2011)}]{Jaffari2}%
  \BibitemOpen
  \bibfield  {author} {\bibinfo {author} {\bibfnamefont {R.}~\bibnamefont
  {Jafari}},\ }\href {\doibase 10.1103/PhysRevB.84.035112} {\bibfield
  {journal} {\bibinfo  {journal} {Phys. Rev. B}\ }\textbf {\bibinfo {volume}
  {84}},\ \bibinfo {pages} {035112} (\bibinfo {year} {2011})}\BibitemShut
  {NoStop}%
\bibitem [{\citenamefont {Sun}\ and\ \citenamefont {Chen}(2009)}]{SunChen}%
  \BibitemOpen
  \bibfield  {author} {\bibinfo {author} {\bibfnamefont {K.-W.}\ \bibnamefont
  {Sun}}\ and\ \bibinfo {author} {\bibfnamefont {Q.-H.}\ \bibnamefont {Chen}},\
  }\href {\doibase 10.1103/PhysRevB.80.174417} {\bibfield  {journal} {\bibinfo
  {journal} {Phys. Rev. B}\ }\textbf {\bibinfo {volume} {80}},\ \bibinfo
  {pages} {174417} (\bibinfo {year} {2009})}\BibitemShut {NoStop}%
\bibitem [{\citenamefont {Wang}\ and\ \citenamefont {Yi}(2010)}]{Wang}%
  \BibitemOpen
  \bibfield  {author} {\bibinfo {author} {\bibfnamefont {L.}~\bibnamefont
  {Wang}}\ and\ \bibinfo {author} {\bibfnamefont {X.}~\bibnamefont {Yi}},\
  }\href@noop {} {\bibfield  {journal} {\bibinfo  {journal} {The European
  Physical Journal D}\ }\textbf {\bibinfo {volume} {57}},\ \bibinfo {pages}
  {281} (\bibinfo {year} {2010})}\BibitemShut {NoStop}%
\bibitem [{\citenamefont {Gu}(2010)}]{Gu}%
  \BibitemOpen
  \bibfield  {author} {\bibinfo {author} {\bibfnamefont {S.}~\bibnamefont
  {Gu}},\ }\href@noop {} {\bibfield  {journal} {\bibinfo  {journal} {Int. J.
  Mod. Phys. B}\ }\textbf {\bibinfo {volume} {24}},\ \bibinfo {pages} {4371}
  (\bibinfo {year} {2010})}\BibitemShut {NoStop}%
\end{thebibliography}%
\end{document}